\documentclass[final,5p,times,twocolumn]{elsarticle}
\usepackage[english]{babel} \usepackage{amsmath} \usepackage{color}
\usepackage{epsfig} \usepackage{graphicx} \usepackage{bm,bbm}
\usepackage{mathtools}

\usepackage{times,amsmath,amssymb,amsfonts}

\journal{Physica A}

\allowdisplaybreaks

\usepackage{hyperref}
\usepackage{amssymb}

\newcommand{\env}[1]{\texttt{#1}}

\newcommand{\be}{\begin{equation}}
\newcommand{\ee}{\end{equation}}

\newcommand{\mediaT}[1]{\left\langle #1 \right\rangle}

\newcommand{\media}[1]{\langle #1 \rangle}

\begin{document}
\begin{frontmatter}

\title{Critical states in Political Trends.\\
How much reliable is a poll on Twitter? A study by means of the Potts Model}

\author[UFSC]{Lucas Nicolao}
\author[UFBA]{Massimo Ostilli}

\address[UFSC]{Departamento de
    F\'isica, Universidade Federal de Santa Catarina, Florian\'opolis, Brazil}
\address[UFBA]{Instituto de
    F\'isica, Universidade Federal da Bahia, Salvador, Brazil}


\begin{abstract}
In recent years, Twitter data related to political trends have tentatively been  used to make predictions (poll) about several electoral events.
Given $q$ candidates for an election and a time-series of Twitts (short messages), 
one can extract the $q$ mean trends and the $q(q+1)/2$ Twitt-to-Twitt correlations, 
and look for the statistical models that reproduce these data. 
On the base of several electoral events and assuming a stationary regime, we find out the following: 
\textit{i)} the maximization of the entropy singles out a microscopic model (single-Twitt-level) that coincides with a $q$-state Potts model 
having suitable couplings and external fields to be determined via an inverse problem from the two sets of data;
\textit{ii)} correlations decay as $1/N_{eff}$, where $N_{eff}$ is a small fraction of the mean number of Twitts;
\textit{iii)} the simplest statistical models that reproduce these correlations are the multinomial distribution (MD), characterized by $q$
external fields, and the mean-field Potts model (MFP), characterized by one coupling;
\textit{iv)} remarkably, this coupling turns out to be always close to its critical value.
This results in a MD or MFP model scenario that discriminates between cases in which polls are reliable and not reliable, respectively. 
More precisely, predictions based on polls should be avoided whenever the data maps to a MFP because anomalous large fluctuations (if $q=2$) or 
sudden jumps (if $q\geq 3$) in the trends might take place as a result of a second-order or a first-order phase transition of the MFP, respectively.
\end{abstract}




\end{frontmatter}

\section{Introduction} 
Social dynamics is a multidisciplinary subject that, especially in the last two decades, has attracted a strong interest
due to an exponential growth of digital data where trends, opinions, and relationships, are nowadays continuously
collected and analyzed. In this context, particularly appealing are the Twitter data related to political orientations. In fact, a vast literature
in which Twitter messages (Twitts) about politic are analyzed, 
has already been produced since Twitter creation in 2006 \cite{GermanTumasjan,Spain2016,India2014,Hauff,Choy,UK2010,Brazil,Brazil-2};
see also \cite{Avello} and references therein.  
In some cases, the analysis of the Twitts collected prior to a particular electoral event in which $q$ candidates compete for the position of president or prime minister, 
have allowed to predict the actual electoral result, while in some others not, even though a fair statistics is lacking \cite{Avello}. 
In these works, the main hypothesis consists in assuming that the set of Twitts collected,  
say, from day -30 till day -1, is representative of the offline will of vote and contains the information about the winner candidate elected on the day 0, the election day.    
Each message's content 
is then analyzed and, according to criteria based on specific dictionaries of key words, the Twitt is to be interpreted in favor (mention),
or possibly against (sentiment), of one of the $q$ candidates.
Any poll, however, is partial and 
predictions must be taken with care, even when the best criteria for interpreting the messages has been used.
The problem with polls is in fact twofold as they are limited in both ``space'' and ``time'':
On one side, samples are finite and the statistics suffers from finite size effects, on the other side, samples look only at specific times and, 
of course, they do not contain any actual information about the future election day. 
If a system 
is somehow at a critical point, the above limitations will dramatically affect the predictions
even under the hypothesis that the system is approximately in a stationary regime. 

In order to justify typical macroscopic patterns observed in social dynamics, several microscopic models have been proposed \cite{Santo}. 
So, for example, on the base of elementary laws
among neighbors, like replication of opinion, competitions, and external noise, different versions of the Voter model \cite{Voter}, as well as 
of the Sznajd dynamics \cite{Sznajd}, have been studied and a large literature produced (see also references in \cite{Santo}). 
Despite important progresses in this direction, especially with respect to certain observed universal properties, like the spread of rumor \cite{Marsili},
and the vote distribution for electoral events characterized by a large number $q$ of candidates (as in the case of legislators and city councilors elections) \cite{Bernardes,Bernardes2,Travesso,Santo2,Crokidakis}, most of such studies have remained mainly qualitative (or spectral quantitative), the major problem being 
the lack of connection between the model-parameters and the actual system-parameters. 
This issue is in fact one of the most serious one for any microscopic approach to social dynamics, for the couplings and the external
fields to which a real social system obeys, if any, depend of an infinite number of different sources out of control,
which results in the impossibility of using the model for making predictions. 

In this paper, we consider a different perspective. We do not aim at finding the most reliable microscopic model from ``first-principles''. 
Rather, we consider effective models which, on the base of a limited number of macroscopic data, 
are able to incorporate the complex source of the collective behavior, and yet are simply enough
to allow for the determination of their actual parameters to be found via an inverse problem, which in turn can reveal surprisingly interesting facts.   
To track the dynamics of a collection of Twitter users posting Twitts about a specific electoral event with $q$ candidates,
in which, at any time $t$, we have access to the ``volume'' (or number of mentions),
\textit{i.e.}, the number of Twitts in favor of a specific candidate and, in some cases 
also to the ``sentiment'', \textit{i.e.}, the number of Twitts in favor minus the number of Twitts against a specific candidate,  
we use a $q$-state Potts model \cite{Wu}. 
The choice of the model is not arbitrary. In fact, as we shall show, 
a generalized $q$-state Potts model with suitable couplings and external fields, in principle time-dependent, is the model that maximizes the entropy
of a set of data constituted by $q$ mean number of mentions and $q(q+1)/2$ mention-to-mention correlations (more precisely, $q$ auto-correlations, and $q(q-1)/2$ two-point correlations). 
In other words, the generalized $q$-state Potts model is maximally random 
under the above $q+q(q+1)/2$ constrains. 
Unfortunately (at the present study), we do not have direct access to the mention-to-mention instantaneous correlations (microscopic level, or single-Twitt level). 
However, under the stationary hypothesis, 
where no impacting news reach the Twitter network, the system can be considered as isolated. 
As a consequence, during a stationary regime, any apparent dynamics can be seen as sole due to stochastic fluctuations originating from a time-independent probability 
distribution and we can derive the macroscopic (or coarse-grained) correlations from the $q$ time series of the mean number of mentions.
Quite interestingly, these macroscopic correlations scale always as $\mathop{O}(1/N_{eff})$, where $N_{eff}$ is an effective number of messages that can also be calculated from the data
and that turns out to be a small fraction of the mean number of messages $N$ 
(where the mean is calculated over the given time interval of observation, ranging from minutes to days).
In turn, this implies that the microscopic correlations scale at most as $\mathop{O}(1/N_{eff})$, \textit{i.e.}, they are effectively mean-field like.
Remarkably, by assuming a mean-field Potts model (MFP) characterized by one single coupling $J/N_{eff}$, we find that, in all the electoral events that we have considered,
the inverse problem is solved for a $J$ close to its critical value. 
However, other mean-field models could work as well. 
In fact, by comparing in detail the structure of the measured macroscopic correlations with the model macroscopic correlations, we see that, 
whereas they are both mean-field like (decay as $1/N_{eff}$),
in some cases the multinomial distribution (MD) turns out to be the best statistical model, while in some others is the MFP.
Note that, in the MD case, microscopic correlations are zero, and macroscopic correlations arise only as a consequence of the constraint that the total number 
of messages is fixed while, in the MFP case, microscopic correlations are present. In both cases, the macroscopic connected correlations functions decay as $1/N_{eff}$, 
but their forms are quite different and allow for a clear distinction between two corresponding scenarios, MD or MFP. Note also that, formally, the MD for $q$ candidates and $N_{eff}$ 
messages corresponds to a $q$-state Potts model with zero coupling and $q$ suitable external fields, while the MFP is characterized by a coupling that scales 
as $1/N_{eff}$ and no external field. Clearly, one could look for other mean-field Potts models with more general features able to interpolate between the MD and the MFP, 
as well as to take into account the scale-free character of the underlying  Twitter network \cite{Neville,TwitterSF}. However, the MD and the MFP, which constitute the two simplest 
and somehow opposite mean-field models, turn out to be able to reproduce well the $q(q+1)/2$ measured macroscopic correlations.
In fact, as we shall show, the use of the effective number of messages $N_{eff}$ 
allows for taking into account replicas of messages, which can be seen as strong-rigid correlations. This aspect makes the MD and MFP models rather non trivial and effective.

The MD \textit{versus} MFP scenario is very interesting.
Our analysis shows that only in the MD case one can use the Twitts as a poll for predictions.
In this case in fact, during an approximate stationary regime, one can simply track the average trends and extrapolate the future election result
by making use of standard regression methods. 
In the case of a MFP model instead, the fact that the model lies in its critical point should alarm the statistician attempting to make predictions about the future electoral result. 
In fact, for $q=2$, where the Potts model is equivalent to the Ising model,
at the critical point the system undergoes a second-order phase transition
and fluctuations are unbounded in the thermodynamic limit. 
For $q\geq 2$, instead, the system undergoes a first-order phase transition where fluctuations are bounded but sudden jumps occur. 
In any case, our analysis shows that, in several electoral events, 
the system tends to settle itself, for over large time intervals, to its critical point, lying then between the two phases: 
a liquid/paramagnetic one where each candidate tends to receive, on average, the same percentage of votes, and a crystal/magnetized one where
one of the candidates breaks the liquid/paramagnetic symmetry gaining a net advantage over the others. The passages between the two phases appear
to be rather unpredictable due to the permanent vicinity to the critical point preventing therefore to make a forecast of the $q$ trends.
At least within the context of democratic elections, this scenario of intrinsic unpredictability seems to be due to the controversial 
spread of political and economic power that takes place among a few symmetrically powerful parties and respective candidates, the strongest
of which rarely overcome $q \sim 10$. Strong candidates are quite able to compete with each other.
Typically, a strong candidate publicly speaking at time say, $t+1$, has a good chance to fade out the impact of the speech 
of her/his strong opponents given at prior times $t, ~t-1,\ldots$.
This process can result in a continuous rank reversal of the trends that, statistically, translates into a system
that remains on the boundary between the two phases. 
In such a scenario of symmetrically strong candidates, the winner of the election is just the result of a random large fluctuation.

The paper is organized as follows.
In Sec. II, we introduce and develop the general inverse problem (IIA) whose formal solution is given in terms of a generalized Potts model having
suitable couplings and external fields (IIB), with important simplifications arising in the homogeneous case (IIC) which, in turn, 
includes the sub-case of zero correlations where the inverse problem is solved by the MD.
In Sec. III, we review the MFP in detail (IIIA).
Although the MFP can be seen as a particular solution of the inverse problem, for pedagogical reasons we prefer to describe it in a separated
section where, in particular, we calculate the connected correlation functions (IIIB).
In Sec. IV, we apply the theory to several Twitter data sets of trends (time series of the mean number of mentions, and also sentiment in a specific case).
In doing this, we first stress the peculiarities of the MD and MFP models (IVA), then we show how to measure the macroscopic correlations in a stationary regime (IVB),
and define the effective number of messages $N_{eff}$ which provides another interesting information on the tendency of the users to copy each other (IVC).
Finally, we apply the theory to several electoral events. At the end, some conclusions are drawn.


\section{Inverse Problem in Social Sciences}
Here we introduce a general unbiased approach that, in principle, allows to derive the actual system-parameters,
although at high computational cost, and also data cost. In fact, it is worth to remind that only certain sets of Twitter data are publicly free.
We first illustrate the procedure in a idealized context where we assume that data are known at the 
user-microscopic level. Next, we shall relax this assumption and make the naive mean-field approximation
in order to cope with the incompleteness of the available data and in view of simplicity.

We assume that there exist $N$ nodes/users/agents/messages and that each agent owns a status variable $\sigma$, the ``spin'', that can take
$q$ possible different values labeled as $\sigma=1,\ldots,q$. 
For example, the state $\sigma$ can represent an opinion on a given topic where only $q$ different answers are possible, or the state can represent the agent' s 
preference over a set of $q$ political candidates. 
The status $\sigma$ must take into account two different tendencies of the spin: a tendency to be influenced by its neighbors,
and a tendency to follow a behavior independently from the others. This latter in turn can be either due to an intrinsic tendency of the user
to follow her/his believes, or can be due to news and events whose sources are external to the set of the $N$ agents.   
In the framework of equilibrium statistical mechanics, the natural way to model such a system consists in using
the $q$-state Potts model, where the couplings $\{J_{i,j}\}$ quantify the tendency of the spins to be influenced by its neighbors,
whereas the external fields $h_i$ quantify the tendency of the spins to follow a behavior independently of the other spins 
(similar ideas have been applied in \cite{Rumor} for the case $q=2$ to analyze the interaction between two communities).
The use of the Potts model in a tentative to describe social interactions is perhaps already available in literature but not abundant for $q>2$.
However, most of the works so far produced in this direction turn out to be quite disconnected from real world phenomena.
The concepts themselves of Hamiltonian, coupling, temperature and external field, which are well established physical entities, 
in social sciences assume a too often abstract meaning and seem to have 
no direct connection with real data. 
In this way, the use of statistical mechanics in social science remains a merely pictorial description of the real phenomena
under study, with no prediction power. This critics applies also to other tools which traditionally do not belong to statistical
mechanics (\textit{e.g.} community detection algorithms, voter-like models, and so on).

We show here that it is instead possible to introduce in social science a rigorous use of statistical mechanics,
where the connection with the real observed data is exact, and where the probabilistic prediction power of statistical mechanics 
is recovered, though at the expenses of high computational costs as well as expensive data. The following probabilistic method is certainly not original 
(see for example the same method applied to graph theory to derive the probability distribution in a canonical ensemble of graphs \cite{Newman}), 
but it is perhaps new in the present context and highlights several points that are important to our aim.

\subsection{The General Problem and its Formal Solution}
Let $\bm{\sigma}$ be the vector of the $N$ spin configurations: $\bm{\sigma}=\left(\sigma_1,\ldots,\sigma_N\right)$.
Let us assume that at any instant $t$ there exists the probability distribution $P\left(\bm{\sigma};t\right)$.
Suppose that all we know about the real phenomena under consideration is a set of $M$ independent averages over the distribution $P\left(\bm{\sigma};t\right)$ of certain observables $Y^{(1)}\left(\bm{\sigma}\right),\ldots,Y^{(M)}\left(\bm{\sigma}\right)$
(in the following we shall omit the dependency on $t$ for brevity):
\begin{eqnarray}
\label{Potts00}
\mediaT{Y^{(l)}}_P=\sum_{\bm{\sigma}}P\left(\bm{\sigma}\right)Y^{(l)}\left(\bm{\sigma}\right),  \quad l=1,\ldots,M.
\end{eqnarray}
We can ``derive'' $P\left(\bm{\sigma}\right)$ as the distribution that maximizes the information entropy 
$S[P]=-\sum_{\bm{\sigma}}P\left(\bm{\sigma}\right)\log\left(P\left(\bm{\sigma}\right)\right)$ under the $M$ constraints (\ref{Potts00}).
In other words, we have to find the distribution $Q\left(\bm{\sigma}\right)$ that maximizes the following functional 
\begin{eqnarray}
\label{Potts001}
&& \widetilde{S}[P',\lambda^{(1)},\ldots,\lambda^{(M)},\mu]=S[P'] \nonumber \\
&& + \sum_{l=1}^M\lambda^{(1)}\left(\mediaT{Y^{(l)}}_{P'}-\mediaT{Y^{(l)}}_P\right)
+\mu\left(\sum_{\bm{\sigma}}P'\left(\bm{\sigma}\right)-1\right),  
\end{eqnarray}
where $P'$ is a trial probability distribution. 
The maximum of $\widetilde{S}[P',\lambda^{(1)},\ldots,\lambda^{(M)},\mu]$ must be found with respect to 
$P'$ and the $M+1$ Lagrangian multipliers, $\lambda^{(1)},\ldots,\lambda^{(M)}$ and $\mu$. The latter Lagrangian multiplier ensures the normalization
of $P'$ while the others ensure that the $M$ constraints (\ref{Potts00}) are satisfied.
The general formal solution of the above problem is the following Gibbs-like distribution 
\begin{eqnarray}
\label{Potts002}
Q\left(\bm{\sigma}\right)=\frac{e^{-H\left(\bm{\sigma}\right)}}{Z},  \qquad Z=\sum_{\bm{\sigma}}e^{-H\left(\bm{\sigma}\right)}
\end{eqnarray}
where
\begin{eqnarray}
\label{Potts003}
H\left(\bm{\sigma}\right)=\sum_{l=1}^M \lambda^{(l)}Y^{(l)}\left(\bm{\sigma}\right), 
\end{eqnarray}
and the $M$ Lagrangian multipliers $\lambda^{(1)},\ldots,\lambda^{(M)}$ solve the following system of Eqs. 
\begin{eqnarray}
\label{Potts004}
&& \mediaT{Y^{(l)}}_P=\mediaT{Y^{(l)}}_{Q} \quad \mathrm{or,~more~explicitly} \nonumber \\
&& \mediaT{Y^{(l)}}_P=\sum_{\bm{\sigma}}\frac{e^{-H\left(\bm{\sigma}\right)}Y^{(l)}\left(\bm{\sigma}\right)}{Z},  \quad l=1,\ldots,M.
\end{eqnarray}
Notice that the normalization condition of $Q$ is automatically satisfied ($\mu$ is absorbed in $Z$).
We stress that, in general, $Q$ does not coincide with the target unknown distribution $P$. In fact, only in the limit of
an infinite number of independent averages we can have the equality between the two distributions (formally, $Q\to P$ for $M\to\infty$).
For $M$ finite, $Q$ is simply the distribution that, under the $M$ constrains (\ref{Potts00}), is maximally random with respect to the distribution $P$. 

More precisely, the use of Shannon' s entropy ensures that: 
\textit{i)} $Q$ is maximally random under the $M$ constrains (there are in principle other functionals that satisfy the same requirement, 
like R\'enyi' s and Tsallis entropy), 
\textit{ii)} if for some data we have, for example, $\mediaT{\sigma_1\sigma_2}_P=\mediaT{\sigma_1}_P\mediaT{\sigma_2}_P$, this means that 
$P(\sigma_1,\sigma_2)=P(\sigma_1)P(\sigma_2)$, a property which is consistent only with an additive (and then extensive) entropy, like the Shannon' s or
the R\'enyi' s entropy, 
\textit{iii)} the chain rule for the conditional probability, $H(\sigma_1|\sigma_2)=H(\sigma_1,\sigma_2)-H(\sigma_2)$ (which is a natural requirement in information theory), is satisfied only 
by the Shannon's entropy.

\subsection{Most general means and correlations}
Let us now consider a system of $N$ agents $\sigma_1,\ldots,\sigma_N$ where, at any instant, $\sigma_i$ can take one over $q$ possible values $1,\ldots,q$.
We assume that, at any instant $t$, we have the following independent data~\footnote{
The sets of data $\overline{x}_{\sigma;i}$ are of course not independent since $\sum_\sigma \overline{x}_{\sigma;i}=1$ for any $i$.
Formally, we can deal with this issue by making use of another Lagrangian multiplier $\mu'$ to be used in Eq. (\ref{Potts001}) that takes into account this 
constrain. However, it is easy to see that $\mu'N$ can be absorbed in $\mu$ with no effect in the final result for $Q(\bm{\sigma})$, 
Eqs. (\ref{Potts002}-\ref{Potts004}) being the same.     
}:
\begin{eqnarray}
\label{Potts1}
\overline{x}_{\sigma;i}\equiv\mediaT{\delta(\sigma_i,\sigma)}_P, \quad i=1,\ldots,N, \quad \sigma=1,\ldots,q, 
\end{eqnarray}
and 
\begin{eqnarray}
\label{Potts2}
C_{\sigma,\sigma';i,j}&\equiv&\mediaT{\delta(\sigma_i,\sigma)\delta(\sigma_j,\sigma')}_P, \nonumber\\
&& i,j=1,\ldots,N, \quad \sigma,\sigma'=1,\ldots,q.
\end{eqnarray}
Notice that in defining the correlation $C_{\sigma,\sigma';i,j}$, we have imposed $i\neq j$ since the information corresponding to $i=j$ is already contained in
the data of $\overline{x}_{\sigma;i}$.  
According to Eqs. (\ref{Potts002})-(\ref{Potts004}), we have the following general Gibbs-Boltzmann form
\begin{eqnarray}
\label{Potts4}
Q(\bm{\sigma})=\frac{e^{- H\left(\bm{\sigma}\right)}}{Z},   \qquad Z=\sum_{\bm{\sigma}}e^{- H\left(\bm{\sigma}\right)}
\end{eqnarray}
where 
\begin{eqnarray}
\label{H}
H(\bm{\sigma})&=&-\sum_i \sum_\sigma h_{\sigma;i} \delta(\sigma_i,\sigma) \nonumber \\
&& -\sum_{i< j} \sum_{\sigma,\sigma'} J_{\sigma,\sigma';i,j}\delta(\sigma_i,\sigma)\delta(\sigma',\sigma_j),
\end{eqnarray}
where $ h_{\sigma;i}$ and $ J_{\sigma,\sigma';i,j}$ are two sets of adimensional parameters, called external fields and couplings, which
must satisfy the constrains (\ref{Potts1}) and (\ref{Potts2}) 
\begin{flalign}
\label{Potts1v}
\sum_{\bm{\sigma}'} \frac{e^{- H\left(\bm{\sigma}'\right)}\delta_{{\sigma}_i',\sigma}}{Z}=\overline{x}_{\sigma;i}, \quad j=1,\ldots,N, \quad \sigma=1,\ldots,q,
\end{flalign}
and 
\begin{eqnarray}
\label{Potts2v}
&&\sum_{\bm{\sigma}'} \frac{e^{- H\left(\bm{\sigma}'\right)}\delta(\sigma_i',\sigma)\delta(\sigma_j',\sigma')}{Z}=C_{\sigma,\sigma';i,j}, \nonumber \\
&&\quad i< j, \quad i,j=1,\ldots,N, \quad \sigma.\sigma'=1,\ldots,q.
\end{eqnarray}
Eqs. (\ref{Potts1v})  and (\ref{Potts2v})  establish our inverse problem:
having the sets of the means $\overline{x}_{\sigma;i}$ and the set of the correlations $C_{\sigma,\sigma';i,j}$, we have the necessary and sufficient
conditions to find, at least numerically, the set of the external fields $ h_{\sigma;i}$, and the set of the
couplings $ J_{\sigma,\sigma';i,j}$. 


\subsection{Homogeneous case}
Above, we have discussed a very general case in which user' s orientation and user-user correlations depend both on the user and on the orientation.
However, in a sufficiently large system, we expect the following homogeneity to take place for the two sets of data:
\begin{eqnarray}
\label{Potts1b}
\overline{x}_{\sigma;i}=\overline{x}_{\sigma}, \qquad i=1,\ldots,N, \qquad \sigma=1,\ldots,q,  
\end{eqnarray}
and 
\begin{flalign}
\label{Potts2b}
C_{\sigma,\sigma';i,j}=C_{\sigma,\sigma'}, ~i< j, \quad i,j=1,\ldots,N, \quad \sigma,\sigma'=1,\ldots,q.
\end{flalign}
The meanings of $\overline{x}_{\sigma}$ and $C_{\sigma,\sigma'}$ are
\begin{eqnarray}
\label{Potts1bb}
\overline{x}_{\sigma}=\mathrm{Prob.}\left(\mathrm{for~a~randomly~chosen~user~to~have~status~}\sigma\right),  
\end{eqnarray}
\begin{eqnarray}
\label{Potts2bb}
C_{\sigma,\sigma'}&=&\mathrm{Prob.}\left(\mathrm{for~two~randomly~chosen~users}\right.\nonumber\\
&&\left.\mathrm{~~~~~~~~~~~to~have~status~}\sigma~\mathrm{and~}\sigma'\right).
\end{eqnarray}
In turn, Eqs. (\ref{Potts1b}) and (\ref{Potts2b}) imply
\begin{eqnarray}
\label{Potts1c}
\overline{x}_{\sigma}=\frac{\mediaT{\sum_i\delta(\sigma_i,\sigma)}_P}{N}=\mediaT{x_{\sigma}}_P, \qquad \sigma=1,\ldots,q, 
\end{eqnarray}
and 
\begin{eqnarray}
\label{Potts2c}
&&C_{\sigma,\sigma'}=\frac{2\mediaT{\sum_{i<j}\delta(\sigma_i,\sigma)\delta(\sigma_j,\sigma')}_P}{N(N-1)}=\nonumber \\
&&\mediaT{x_{\sigma}x_{\sigma'}}_P-\frac{\delta_{\sigma,\sigma'}}{N}\mediaT{x_{\sigma}}_P,
\qquad \sigma,\sigma'=1,\ldots,q,
\end{eqnarray}
where we have used $N-1\simeq N$, and introduced the ``frequency'' random variable
\begin{eqnarray}
\label{g13}
x_{\sigma}=\frac{1}{N}\sum_i \delta(\sigma_i,\sigma).
\end{eqnarray}  

Eqs. (\ref{Potts1c})  and (\ref{Potts2c}) are important because they are expressed in terms of the random variables $x_\sigma$'s which constitute the only information available to us. 

The homogeneous assumptions (\ref{Potts1b})  and (\ref{Potts2b}) imply that the Hamiltonian takes the following form
\begin{eqnarray}
\label{Hb}
  H(\bm{\sigma})&=&-\sum_{\sigma,\sigma'} J_{\sigma,\sigma'}\sum_{i<j}\delta(\sigma_i,\sigma)\delta(\sigma_j,\sigma')\nonumber \\
&& -\sum_\sigma h_{\sigma} \sum_i\delta(\sigma_i,\sigma),
\end{eqnarray}
where the sets of external fields and couplings must satisfy
\begin{eqnarray}
\label{Potts1vb}
\sum_{\bm{\sigma}'} \frac{e^{- H\left(\bm{\sigma}'\right)}\delta_{{\sigma}_1',\sigma}}{Z}=\overline{x}_{\sigma}, \qquad \sigma=1,\ldots,q,
\end{eqnarray}
and 
\begin{eqnarray}
\label{Potts2vb}
\sum_{\bm{\sigma}''} \frac{e^{- H\left(\bm{\sigma}''\right)}\delta_{\sigma_1'',\sigma}\delta_{\sigma_2'',\sigma'}}{Z}=C_{\sigma,\sigma'}, 
\qquad \sigma,\sigma'=1,\ldots q.
\end{eqnarray}

It is understood that, in the homogeneous case, we are free to measure the correlations by means of Eqs. (\ref{Potts2}) to be used in the inverse
problem via Eq. (\ref{Potts2vb}), or by means of $\mediaT{x_{\sigma}x_{\sigma'}}_P$ via 
\begin{eqnarray}
\label{Potts2vc}
\sum_{\bm{\sigma}''} \frac{e^{- H\left(\bm{\sigma}''\right)} x_{\sigma}x_{\sigma'}}{Z}=\mediaT{x_{\sigma}x_{\sigma'}}_P, 
\qquad \sigma,\sigma'=1,\ldots q.
\end{eqnarray}
and similarly for the connected correlation functions.

\subsection*{Homogeneous uncorrelated case - Multinomial Distribution} 
This case applies when the $\sigma_i$'s are independent and uniformly distributed random variables.
As a consequence, trivial correlations exist only for the frequency random variable $x_\sigma$'s, Eqs. (\ref{g13}).
In fact, due to the constraint $\sum_\sigma x_\sigma=1$, it is easy to see that 
\begin{eqnarray}
\label{Uncorr4H}
\media{x_{\sigma}x_{\sigma'}}-\media{x_{\sigma}}\media{x_{\sigma'}}=
\frac{1}{N}\left[\overline{x}_\sigma\delta_{\sigma,\sigma'}-\overline{x}_\sigma\overline{x}_{\sigma'}\right],
\end{eqnarray}  
and the probability distribution for the $x_\sigma$'s, $Q(\bm{\sigma})$, turns to be the multinomial distribution 
\footnote{
with an abuse of terminology we shall say that the probability distributions for both the $x_\sigma$'s and the $N_\sigma$'s are MD's.}
\begin{eqnarray}
\label{Uncorr2H}
\widetilde{Q}(\bm{N}_\sigma)=N!\prod_\sigma \frac{\overline{x}^{N_\sigma}_{\sigma}}{N_\sigma!},
\end{eqnarray}
where $N_\sigma=\sum_i \delta(\sigma,\sigma_i)$, proportional to the random variable $x_\sigma$, is a new random variable counting how many users have status $\sigma$,
in terms of which, the connected correlation functions are
\begin{eqnarray}
\label{Uncorr3H}
\media{N_{\sigma}N_{\sigma'}}-\media{N_{\sigma}}\media{N_{\sigma'}}=
N\left[\overline{x}_\sigma\delta_{\sigma,\sigma'}-\overline{x}_\sigma\overline{x}_{\sigma'}\right].
\end{eqnarray}
It is possible to derive the multinomial distribution (\ref{Uncorr3H}) by using the previous approach as follows.
From Eq. (\ref{Hb}) we see that, since the $\sigma_i$'s have no correlations, we can look for solutions $Q(\bm{\sigma})$ in which $J_{\sigma.\sigma'}=0$ and we find
\begin{eqnarray}
\label{Uncorr5H}
Q(\bm{\sigma})=\prod_{i=1} \frac{e^{-h_{\sigma_i}}}{\sum_\sigma e^{-h_{\sigma}}}.
\end{eqnarray}
The fields $h_\sigma$'s can be expressed in terms of the means $\overline{x}_\sigma$'s by applying Eq. (\ref{Potts1c}), which provides the following formal system of $q$ equations
\begin{eqnarray}
\label{Uncorr6H}
\frac{e^{-h_ \sigma  }}{\sum_{\sigma'} e^{-h_{\sigma'}}}=\overline{x}_\sigma, \quad \sigma=1,\ldots,q.
\end{eqnarray}
On plugging Eq. (\ref{Uncorr6H}) into Eq. (\ref{Uncorr5H}) we obtain
\begin{eqnarray}
\label{Uncorr7H}
Q(\bm{\sigma})=\prod_{i=1}\overline{x}_{\sigma_i}=\prod_\sigma\overline{x}^{N_\sigma}_{\sigma}.
\end{eqnarray}
On noting that the last expression depends only on the multiplicity random variables $N_\sigma$'s, we can finally obtain the distribution for the latter,
$\widetilde{Q}(\bm{N}_\sigma)$, by multiplying $Q(\bm{\sigma})$ by the number of ways in which we can arrange the vector $\{N_\sigma\}$ among $N$ spins, $N!/(\prod_\sigma N_\sigma!)$, and the result is
the multinomial distribution (\ref{Uncorr2H}).  
Later on, we shall make use of Eqs. (\ref{Uncorr4H}) for a crucial benchmark that distinguishes between interacting and non interacting models. 

\section{The mean-field Potts model}
When a set of data $\{\overline{x}_{\sigma}\}$ displays a symmetry (or some approximate symmetry) among $q$ or $q-1$ of its components, the distribution $Q(\bm{\sigma})$
in which $H(\bm{\sigma})$ corresponds to the mean-field Potts model, turns out to be a good candidate for solving the inverse problem of the previous Section.
Below, we review the traditional $q$-state mean-field Potts model
emphasizing some points not often stressed in literature. We shall evaluate, in particular, the connected correlation functions for $N$ finite
(to the best of our knowledge, this constitutes a new result). A posteriori, we can say that the latter allow to rigorously establish under which hypotheses on the data
the inverse problem is solved by the distribution $Q(\bm{\sigma})$ associated to the mean-field Potts model. Indeed, since the complete analytical solution (direct problem)
of the mean-field Potts model is rather non trivial, it is pedagogically much more convenient to devote an entire Section to the solution of the direct problem.

\subsection{Uniform coupling and homogeneous fields}
The mean-field Potts model with a uniform coupling and homogeneous fields is defined through the following Hamiltonian built on the
fully connected (or complete) graph
\begin{eqnarray}
\label{HgPotts}
H=-\frac{J}{N}\sum_{i<j}\delta(\sigma_i,\sigma_j)-\sum_\sigma h_\sigma\sum_i\delta(\sigma,\sigma_i),
\end{eqnarray}
where $\delta(\sigma,\sigma')$ is the Kronecker delta function.
Let us rewrite $H$ as (up to terms negligible for $N\to\infty$)
\begin{eqnarray}
\label{HgPotts00}
H=-\frac{J}{N}\sum_\sigma \left[\sum_{i}\delta(\sigma_i,\sigma)\right]^2-\sum_\sigma h_\sigma\sum_i\delta(\sigma,\sigma_i).
\end{eqnarray} 
From Eq. (\ref{HgPotts00}) we see that, if $J>0$, by introducing $q$ independent Gaussian variables, $x_\sigma$, $\sigma=1,\ldots,q$, we can
evaluate the partition function, $Z= \sum_{\{\sigma_i\}}\exp(- H(\{\sigma_i\}))$, as
\begin{eqnarray}
\label{Z1Potts0}
Z\propto \int \prod_{\sigma=1}^{q} d x_\sigma ~ e^{-N\left[\sum_\sigma \frac{ J x_\sigma^2}{2}
-\log\left(\sum_\sigma e^{ J x_{\sigma}+ h_\sigma}\right)
\right]}.
\end{eqnarray}  
From Eq. (\ref{Z1Potts0}), for $N\to\infty$, 
we get immediately
the following system of equations for the saddle point $\{x_\sigma^{\mathrm{sp}}\}$
\begin{eqnarray}
\label{Potts0t}
x^{\mathrm{sp}}_{\sigma}=\frac{e^{ J x^{\mathrm{sp}}_\sigma+ h_\sigma}}{\sum_{\sigma'}e^{ J x^{\mathrm{sp}}_{\sigma'}+ h_{\sigma'}}}, \quad \sigma=1,\ldots,q,
\end{eqnarray}  
while the free energy density $f$ is given by
\begin{eqnarray}
\label{Pottsf0t}
 f=-\log\left(\sum_\sigma e^{ J x^{\mathrm{sp}}_{\sigma}+ h_\sigma}\right)+\sum_\sigma \frac{ J (x^{\mathrm{sp}}_\sigma)^2}{2},
\end{eqnarray}  
to be evaluated in correspondence of the solution of the 
system (\ref{Potts0t}).
For each $\sigma$, $x^{\mathrm{sp}}_\sigma$ coincides
with the thermal average of $\sum_i\delta(\sigma_i,\sigma)/N$,
\textit{i.e.}, the probability to find any spin in the state $\sigma$.
For $\{h_\sigma=0\}$ Eqs. (\ref{Potts0t}) are symmetric under permutation of the components
$(x^{\mathrm{sp}}_1,\ldots,x^{\mathrm{sp}}_q)$.
Moreover, for $\{h_\sigma=0\}$ all the possible solutions of Eqs. (\ref{Potts0t}) can be found 
by setting $q-1$ components equal to each other and solving one
single equation.
If $(i_1,\ldots,i_q)$ is any permutation of $(1,\ldots,q)$, then we set
\begin{eqnarray}
\label{UA}
x^{\mathrm{sp}}_{i_1}=x, \quad x^{\mathrm{sp}}_{i_j}=y, \quad j=2,\ldots,q
\end{eqnarray}  
where $y=(1-x)/(q-1)$, and $x$ satisfies the equation 
\begin{eqnarray}
\label{UA1}
x=\frac{1}{1+(q-1)\exp\left[\frac{ J(1-qx)}{q-1}\right]}.
\end{eqnarray}  
Eqs. (\ref{Pottsf0t})-(\ref{UA1}) give rise to a well known phase transition scenario \cite{Wu}:
a second-order mean-field Ising phase transition sets up only for $q=2$ at the critical coupling $J_c=2$, while for any 
$q\geq 3$ there is a first-order phase transition at the following critical value
\begin{eqnarray}
\label{tc}
J_c = \frac{2(q-1)}{q-2}\log(q-1).
\end{eqnarray}  

\subsection{Covariances and correlations in the case of uniform coupling and homogeneous fields}
In order to calculate correlations in the case of homogeneous fields, we have
to first generalize the previous calculation to include the external fields site- and status- dependent: $\{h_{\sigma,i}\}$.
The generalized Hamiltonian is
\begin{eqnarray}
\label{HgPotts1}
H=-\frac{J}{N}\sum_{i<j}\delta(\sigma_i,\sigma_j)-\sum_\sigma \sum_ih_{\sigma,i}\delta(\sigma,\sigma_i),
\end{eqnarray}
and the generalized partition function is
\begin{eqnarray}
\label{Z1Potts1}
Z(\{h_{\sigma,i}\})&\propto& \int \prod_{\sigma=1}^{q} d x_\sigma ~ e^{-N\left[\sum_\sigma \frac{ J x_\sigma^2}{2}
-\frac{1}{N}\sum_i\log\left(\sum_\sigma e^{ J x_{\sigma}+ h_{\sigma,i}}\right)
\right]} \nonumber \\
&&\propto \int \prod_{\sigma=1}^{q} d x_\sigma ~ e^{-Ng\left(\{x_\sigma;h_{\sigma,i}\}\right)}, 
\end{eqnarray}  
where
\begin{eqnarray}
\label{g}
g\left(\{x_\sigma;h_{\sigma,i}\}\right)=\sum_\sigma \frac{ J x_\sigma^2}{2}-\frac{1}{N}\sum_i\log\left(\sum_\sigma e^{ J x_{\sigma}+ h_{\sigma,i}}\right).
\end{eqnarray}  
We have
\begin{eqnarray}
\label{g1}
\frac{\partial g\left(\{x_\sigma;h_{\sigma,i}\}\right)}{\partial x_\sigma}= J x_\sigma- J\frac{1}{N}\sum_i \frac{e^{ J x_{\sigma}^\mathrm{sp}+ h_{\sigma,i}}}{\sum_{\sigma'}e^{ J x_{\sigma'}^\mathrm{sp}+ h_{\sigma',i}}}.
\end{eqnarray}  
The saddle point $\{x_\sigma^{\mathrm{sp}}\}$ of the integral in Eq. (\ref{Z1Potts1}) corresponds to the system of $q\times q$ Eqs.
\begin{eqnarray}
\label{g2}
x_\sigma^{\mathrm{sp}}=\frac{1}{N}\sum_i \frac{e^{ J x_{\sigma}^\mathrm{sp}+ h_{\sigma,i}}}{\sum_{\sigma'}e^{ J x_{\sigma'}^\mathrm{sp}+ h_{\sigma',i}}}, 
\quad \sigma=1,\ldots,q
\end{eqnarray}  
and the function $g$ evaluated at the saddle point, $g^\mathrm{sp}$, is
\begin{flalign}
\label{g3}
g^\mathrm{sp}\left(\{h_{\sigma,i}\}\right)=\left[
\sum_\sigma \frac{ J x_\sigma^2}{2}-\frac{1}{N}\sum_i\log\left(\sum_\sigma e^{ J x_{\sigma}+ h_{\sigma,i}}\right)\right]_{\{x_\sigma=x_{\sigma}^{\mathrm{sp}}\}},
\end{flalign}  
which provides the partition function as
\begin{eqnarray}
\label{Z1Potts3}
Z(\{h_{\sigma,i}\})\propto ~ e^{-Ng^\mathrm{sp}\left(\{h_{\sigma,i}\}\right)}.
\end{eqnarray}  
The one-point and two-point connected correlation functions are related to $g^\mathrm{sp}$ via
\begin{eqnarray}
\label{g4}
\mediaT{\delta(\sigma_i,\sigma)}=-N\frac{\partial g^\mathrm{sp}\left(\{h_{\sigma,i}\}\right)}{\partial h_{\sigma,i}},
\end{eqnarray}  
and
\begin{eqnarray}
\label{g5}
&& \mediaT{\delta(\sigma_i,\sigma)\delta(\sigma_j,\sigma')}-\mediaT{\delta(\sigma_i,\sigma)}\mediaT{\delta(\sigma_j,\sigma')}=
\nonumber \\
&&-N\frac{\partial^2 g^\mathrm{sp}\left(\{h_{\sigma,i}\}\right)}{\partial h_{\sigma,i}\partial h_{\sigma',j}}, \quad i\neq j.
\end{eqnarray}  
By using Eqs. (\ref{g2}), (\ref{g4}) and (\ref{g5}), we get
\begin{eqnarray}
\label{g6}
\mediaT{\delta(\sigma_i,\sigma')}=\frac{e^{ J x_{\sigma}^\mathrm{sp}+ h_{\sigma,i}}}{\sum_{\sigma'}e^{ J x_{\sigma'}^\mathrm{sp}+ h_{\sigma',i}}}
\end{eqnarray}  
and
\begin{flalign}
\label{g7}
&\mediaT{\delta(\sigma_i,\sigma)\delta(\sigma_j,\sigma')}-\mediaT{\delta(\sigma_i,\sigma)}\mediaT{\delta(\sigma_j,\sigma')}=&&\nonumber \\ 
&\frac{e^{ J x_{\sigma}^\mathrm{sp}+ h_{\sigma,i}}}{\sum_{\sigma''}e^{ J x_{\sigma''}^\mathrm{sp}+ h_{\sigma'',i}}}J\frac{\partial x^{\mathrm{sp}}_\sigma}{\partial h_{\sigma',j}}&&  \nonumber \\
&-\frac{e^{ J x_{\sigma}^\mathrm{sp}+ h_{\sigma,i}}}{\left(\sum_{\sigma''}e^{ J x_{\sigma''}^\mathrm{sp}+ h_{\sigma'',i}}\right)^2}\sum_{\sigma''}e^{ J x_{\sigma''}^\mathrm{sp}+ h_{\sigma'',i}}J 
\frac{\partial x^{\mathrm{sp}}_{\sigma''}}{\partial h_{\sigma',j}}, \quad i\neq j. &&
\end{flalign}  
The partial derivatives of $x^{\mathrm{sp}}_\sigma$ can be obtained by solving the following system 
derived from the saddle point Eqs. (\ref{g2})
\begin{eqnarray}
\label{g8}
&&\frac{\partial x^{\mathrm{sp}}_\sigma}{\partial h_{\sigma',i}}=
Jx^{\mathrm{sp}}_\sigma\frac{\partial x^{\mathrm{sp}}_\sigma}{\partial h_{\sigma',i}}+
\frac{1}{N}\frac{e^{ J x_{\sigma'}^\mathrm{sp}+ h_{\sigma',i}}}{\sum_{\sigma''}e^{ J x_{\sigma''}^\mathrm{sp}+ h_{\sigma'',i}}}\delta_{\sigma,\sigma'} \nonumber \\&-&
\frac{1}{N}\sum_j\frac{e^{ J x_{\sigma}^\mathrm{sp}+ h_{\sigma,j}}}{\left(\sum_{\sigma''}e^{ J x_{\sigma''}^\mathrm{sp}+ h_{\sigma'',j}}\right)^2}
\sum_{\sigma''}e^{ J x_{\sigma''}^\mathrm{sp}+ h_{\sigma'',j}}J\frac{\partial x^{\mathrm{sp}}_{\sigma''}}{\partial h_{\sigma',i}}\nonumber \\
&& -
\frac{1}{N}\frac{e^{ J x_{\sigma}^\mathrm{sp}+ h_{\sigma,i}} ~e^{ J x_{\sigma'}^\mathrm{sp}+ h_{\sigma',i}}}{\left(\sum_{\sigma''}e^{ J x_{\sigma''}^\mathrm{sp}+ h_{\sigma'',i}}\right)^2},
\end{eqnarray}  
from which we see that
\begin{eqnarray}
\label{g9}
\frac{\partial x^{\mathrm{sp}}_\sigma}{\partial h_{\sigma',i}}=\mathrm{O}\left(\frac{1}{N}\right) \quad \mathrm{and} \quad 
\frac{\partial^2 x^{\mathrm{sp}}_\sigma}{\partial h_{\sigma',i}\partial h_{\sigma'',j}}=\mathrm{O}\left(\frac{1}{N^2}\right).
\end{eqnarray}  
As a consequence, as expected, we get
\begin{eqnarray}
\label{g10}
\mediaT{\delta(\sigma_i,\sigma)}=\mathrm{O}\left(1\right)
\end{eqnarray}  
and
\begin{flalign}
\label{g11}
\mediaT{\delta(\sigma_i,\sigma)\delta(\sigma_j,\sigma')}-\mediaT{\delta(\sigma_i,\sigma)}\mediaT{\delta(\sigma_j,\sigma')}=\mathrm{O}\left(\frac{1}{N}\right), \quad i\neq j.
\end{flalign}  
Notice that the last Eq. holds only for $i\neq j$, while for $i=j$
\begin{flalign} 
\label{g12}
&\mediaT{\delta(\sigma_i,\sigma)\delta(\sigma_i,\sigma')}-\mediaT{\delta(\sigma_i,\sigma)}\mediaT{\delta(\sigma_i,\sigma')}=&\nonumber \\
&\mediaT{\delta(\sigma_i,\sigma)}\delta(\sigma,\sigma')-\mediaT{\delta(\sigma_i,\sigma)}\mediaT{\delta(\sigma_i,\sigma')}=
\mathrm{O}\left(1\right).&
\end{flalign}
From Eqs. (\ref{g6})-(\ref{g8}) (valid for $N$ large), and by using the ``frequency'' random variable (\ref{g13}), we get immediately
\begin{eqnarray}
\label{g14}
\media{x_{\sigma}}=x^{\mathrm{sp}}_\sigma.
\end{eqnarray}  
The exact evaluation of Eq. (\ref{g7}) is not trivial. For site-independent external fields, $\{h_{\sigma,i}\equiv h_{\sigma}\}$, 
Eq. (\ref{g8}) becomes
\begin{eqnarray}
\label{g15}
\left[\frac{\partial x^{\mathrm{sp}}_\sigma}{\partial h_{\sigma',i}}\right]_{\{h_{\sigma,i}\equiv h_{\sigma}\}} &=&
\left[Jx^{\mathrm{sp}}_\sigma\frac{\partial x^{\mathrm{sp}}_\sigma}{\partial h_{\sigma'}}
-Jx^{\mathrm{sp}}_{\sigma}\sum_{\sigma''}x^{\mathrm{sp}}_{\sigma''}\frac{\partial x^{\mathrm{sp}}_{\sigma''}}{\partial h_{\sigma'}}\right. \nonumber \\
&& \left. +\frac{1}{N}\left(x^{\mathrm{sp}}_{\sigma}\delta_{\sigma,\sigma'}-x^{\mathrm{sp}}_{\sigma}x^{\mathrm{sp}}_{\sigma'}\right)\right]_{\{h_{\sigma,i}\equiv h_{\sigma}\}},
\end{eqnarray}  
which, plugged into Eq. (\ref{g7}) gives
\begin{flalign}
\label{g16}
&\left[\mediaT{\delta(\sigma_i,\sigma)\delta(\sigma_j,\sigma')}-
\mediaT{\delta(\sigma_i,\sigma)}\mediaT{\delta(\sigma_j,\sigma')}\right]_{\{h_{\sigma,i}\equiv h_{\sigma}\}}=&&\nonumber \\
&\left[\frac{\partial x^{\mathrm{sp}}_\sigma}{\partial h_{\sigma',i}}-
\frac{1}{N}\left(x^{\mathrm{sp}}_{\sigma}\delta_{\sigma,\sigma'}-x^{\mathrm{sp}}_{\sigma}x^{\mathrm{sp}}_{\sigma'}\right)
\right]_{\{h_{\sigma,i}\equiv h_{\sigma}\}}, \quad i\neq j,&&
\end{flalign}  
and then, by also using Eq. (\ref{g12}) and definition (\ref{g13}), we obtain
\begin{eqnarray}
\label{g17}
\left[\media{x_{\sigma}x_{\sigma'}}-\media{x_{\sigma}}\media{x_{\sigma'}}\right]_{\{h_{\sigma,i}=h_\sigma\}}=
\left[\frac{\partial x^{\mathrm{sp}}_\sigma}{\partial h_{\sigma',i}}\right]_{\{h_{\sigma,i}\equiv h_{\sigma}\}}.
\end{eqnarray}  
An easy case occurs for $J\leq J_c$ and $h_\sigma=0$, where $J_c$ is given by Eq. (\ref{tc}) and $x_\sigma\equiv 1/q$, by using which in Eq. (\ref{g15}), together with $\sum_\sigma x_\sigma=1$, we find (formally)
\begin{eqnarray}
\label{g18}
&&\left[\media{x_{\sigma}x_{\sigma'}}-\media{x_{\sigma}}\media{x_{\sigma'}}\right]_{\{h_{\sigma,i}=h_\sigma\}}=\nonumber \\
&&\frac{1}{N\left(1-\frac{J}{q}\right)}\left[x^{\mathrm{sp}}_{\sigma}\delta_{\sigma,\sigma'}-x^{\mathrm{sp}}_{\sigma}x^{\mathrm{sp}}_{\sigma'}\right]_{\{h_{\sigma,i}\equiv h_{\sigma}=0\}}.
\end{eqnarray}  
It is easy to check that, if $J\leq J_c$, for $q>2$ it is $1-J/q> 0$, so that, in the paramagnetic phase, up to a positive constant,
the correlations of the mean-field Potts model have the same formal structure of the correlations of a multinomial distribution where $\overline{x}_\sigma=1/q$ for any $\sigma$,
see Eqs. (\ref{Uncorr4H}).

In general, Eq. (\ref{g15}) can be put in a $q\times q$ matrix form by
\begin{eqnarray}
\label{g19}
\bm{A}=\frac{1}{N}\left[\bm{I}-J\left(\bm{x}-\bm{x}\cdot\bm{x}^T\right)\right]^{-1}\left(\bm{x}-\bm{x}\cdot\bm{x}^T\right),
\end{eqnarray}  
where $\bm{I}$ is the $q\times q$ identify matrix, 
\begin{eqnarray}
\label{g20}
A_{\sigma,\sigma'}=\left[\frac{\partial x^{\mathrm{sp}}_\sigma}{\partial h_{\sigma',i}}\right]_{\{h_{\sigma,i}\equiv h_{\sigma}\}},
\end{eqnarray}  
\begin{eqnarray}
\label{g21}
x_{\sigma,\sigma'}=\left[x^{\mathrm{sp}}_\sigma\right]_{\{h_{\sigma,i}\equiv h_{\sigma}\}}\delta_{\sigma,\sigma'},
\end{eqnarray}  
and $\bm{x}\cdot\bm{x}^T$ stands for the dyadic product between the vector $(x_1,\ldots,x_q)^T$ and itself.

\section{Application to Twitter data for political trends}
Twitter is a micro blogging environment where users post small messages, or Twitts, depicting their
likes and dislike towards a certain topic, e.g. candidates to the next political elections.
In this Section, we apply the previous general approach to Twitter data related to political trends.

We have considered a number of electoral events and used
Twitter data collected and published in papers and/or websites \cite{Brazil,Brazil-2,UK2010,Spain2016,India2014}.
In the Refs. \cite{Brazil,Brazil-2,UK2010,Spain2016,India2014}, the following general scheme is used:
For each electoral event there are $q$ main candidates, each candidate is represented by a label $\sigma=1,\ldots,q$, 
and the activity of a large number of users (order $10^5$ or more) is followed for $T$ days (or $T$ minutes) prior or after the election day (or election minute), 
or partially prior and after. 
The days (minutes) of observation are represented by the integer $t=1,\ldots,T$, while the corresponding collected number of messages (order $10^6$ or more) is represented by $N(t)$.   
At each $t$, the activity of the $i$-th message, $i=1,\ldots,N(t)$, is then analyzed 
by looking for key words providing the political preference of the user toward one specific candidate among the $q$ ones:
$\sigma_i(t)$.
The $q$ percentages $x_\sigma(t)=\sum_i\delta_{\sigma_i(t),\sigma}/N(t)$, or normalized mean number of mentions, as in Eq. (\ref{Potts1c}),
are then tracked till the final time $T$.
Taking into account that $\sum_\sigma x_\sigma(t)=1$, these data constitute a set of $(q-1)T$ independent information.
Consistently with the general approach derived in Sec. II, the statistical model to be used for these data should correspond to the generalized Potts model via Eq. (\ref{Potts4}) where, besides being totally general, the couplings and external fields are also time dependent.
Notice that, whereas in Sec. II, $N$ represents the number of users (or agents), here $N$ (at each time $t$) refers to the number of messages.
We stress that, from a probabilistic point of view, the role of $N$ in Sec. II, and the role of $N$ here, are the same.
Here, the advantage in working with the number of messages rather than the number of users, relies on the fact that each user, in general, at each time $t$ is free to post an arbitrary
number of messages (whereas in Sec. II each user/agent was supposed to take a unique status at each $t$). 
It is then clear that $x_\sigma(t)$ in Twitter, in general, does not represent the mean offline preference toward the $\sigma$ candidate since
a user of Twitter could repeat her/his preference many times while an offline person can express her/his vote just once. On the other hand, it is also possible that each Twitter user
repeats her/his preference the same number of times on average. In such a case, the normalized mean number of mentions $x_\sigma(t)$ would be representative of the offline preference.
At any rate, the main aim of this work is to look for the models that best reproduce the observed Twitter data, independently of the actual offline will of vote.

\subsection{Statistical Models: Multinomial Distribution (MD) \textit{vs} Mean-field Potts (MFP)}
Unfortunately (currently) the percentages $x_\sigma(t)$ constitute the only data available to us. In particular, we have no direct access to
the single message activity $\sigma_i(t)$, nor to the instantaneous correlation functions. As a consequence, even under the very mild assumption of homogeneity,
where the Potts model takes the form (\ref{Hb}), we have no way to determine the involved general couplings $J_{\sigma,\sigma'}$ and external fields $h_\sigma$. 
As mentioned in the Introduction, we overcome this impasse by assuming a stationary regime and 
by using the empirical observation that the macroscopic correlations (see below) decay inversely with an effective number of messages $N_{eff}$.
The stationary hypothesis allows us to measure the macroscopic correlations by using the $x_\sigma(t)$ data, while the weak correlations    
lead to assume one of the two possible statistical models: MD and MFP.
By summarizing the two possible statistical models are characterized as follows.

\begin{itemize}
\item Multinomial distribution (MD): the $\sigma_i$'s are independent random variables taking $q$ possible values according to $q$ arbitrary probabilities $\overline{x}_\sigma$ normalized to 1; 
as a consequence, weak (order $1/N_{eff}$), and trivial (due to the constraint $\sum_\sigma x_\sigma=1$) macroscopic correlations, 
are present only in the frequency random variables $x_\sigma$'s (not in the $\sigma_i$'s ), as given
by Eqs. (\ref{Uncorr4H}). This model reproduces a urn of $N_{eff}$ elements taking each a color among $q$ possible ones 
biased by the probabilities $\overline{x}_\sigma$. As it will be made clear soon, the MD model works well also when the messages are strongly correlated via replication.

\item Mean-field Potts model (MFP): the $\sigma_i$'s are weakly (order $1/N_{eff}$) correlated random variables taking $q$ possible values and
the only parameter is a uniform coupling $J/N_{eff}$. 
This model reproduces a kind of self-organized system where the users have not arbitrary preferences for the latter are determined by the coupling $J$.
In particular, there exists a critical value below which the system is symmetric with
$\overline{x}_\sigma\equiv 1/q$, and above which the system has a winner and $q-1$ equal losers, as determined by Eqs. (\ref{UA}-\ref{UA1}). 
The non trivial correlations are given by Eqs. (\ref{g17}) to be solved in combination with Eqs. (\ref{g19}-\ref{g21}).
Also in this case, the MFP model works well even if the messages are strongly correlated via replication.
\end{itemize}

The above two models are both mean-field like and constitute two opposite limits. We have considered also some intermediate cases involving both a coupling 
and external fields for which Eqs. (\ref{Potts0t}) apply. 
However, as it will be evident in the next examples, such attempts fail as the coupling and external fields strongly fluctuate, contradicting
the stationary hypothesis. We shall come back on this issues in the Conclusions.

\subsection{Sampled Means and Correlation functions - Stationary regimes}
For both the models, MD and MFP, the correlation functions of the random variables $x_{\sigma}$'s (model macroscopic correlation functions) can be expressed
in terms of the sole means $\overline{x}_\sigma$'s. On the other hand, within the stationary assumption 
(which means that the $x_\sigma(t)$'s originate from a time independent probability distribution)
we can also measure the macroscopic correlations from the $x_\sigma(t)$ time series. 
In fact, such a condition certainly applies
when no news arrives and the users keep or change their status only due to closed interactions within the Twitter community.
Under such circumstances, we can evaluate the sampled means and the correlation functions as
\begin{eqnarray}
\label{MeanData}
\mediaT{x_\sigma}_{P}=\overline{x}_\sigma=\frac{1}{T}\sum_{t=1}^{T}x_\sigma(t),
\end{eqnarray}  
and
\begin{eqnarray}
\label{CorrData}
\mediaT{x_\sigma x_{\sigma'}}_{P}=\frac{1}{T}\sum_{t=1}^{T}x_\sigma(t)x_{\sigma'}(t).
\end{eqnarray}  
However, for properly taking into account the fact that the number of messages $N(t)$ strongly fluctuates, we shall rather use the following weighted means and correlations
\begin{eqnarray}
\label{MeanDataw}
\mediaT{x_\sigma}_{P}=\overline{x}_\sigma=\frac{\sum_{t=1}^{T}N(t)x_\sigma(t)}{\sum_{t=1}^{T}N(t)},
\end{eqnarray}  
and
\begin{eqnarray}
\label{CorrDataw}
\mediaT{x_\sigma x_{\sigma'}}_{P}=\frac{\sum_{t=1}^{T}N(t)x_\sigma(t)x_{\sigma'}(t)}{\sum_{t=1}^{T}N(t)}.
\end{eqnarray}  
Eqs. (\ref{MeanDataw}) and (\ref{CorrDataw}) prevail over Eqs. (\ref{MeanData}) and (\ref{CorrData}) because the larger is $N(t)$ the larger is the corresponding statistics.
Concerning the MD, Eqs. (\ref{MeanDataw}) are to be used in Eqs. (\ref{Uncorr4H}) for the model correlations.
Concerning the MFP, the solution of the inverse problem generates a time series of couplings $J(t)$ from which we can evaluate the weighted mean via
\begin{eqnarray}
\label{MeanJ}
\mediaT{J}_{P}=\frac{\sum_{t=1}^{T}N(t)J(t)}{\sum_{t=1}^{T}N(t)},
\end{eqnarray}  
which in turn, in combination with Eqs. (\ref{MeanDataw}), is to be used in Eqs. (\ref{g17}) for the model correlations.

\subsection{The mean number and the effective number of Twitts ($N$ and $N_{eff}$)}
In \cite{Brazil,Brazil-2,UK2010,Spain2016,India2014}, the time series $N(t)$ of the number of Twitts followed in each electoral event are provided.
From each series one can in particular calculate the weighted mean number of messages over the period $T$ as follows
\begin{eqnarray}
\label{Nw}
N=\frac{\sum_{t=1}^{T}N(t)N(t)}{\sum_{t=1}^{T}N(t)}.
\end{eqnarray}  
However, this value of $N$, in general, does not correspond to the number entering the formulas of the MD and MFP models.
In fact, these models are effective and an effective number of messages $N_{eff}$ is required. Let us first consider the MD model. 
This model turns out to be exact if each message is independent from the others, however,
this model turns out to be exact even when each message replicates the behavior of some other messages. The groups of messages having the same
behavior will be distributed according to a MD characterized by a suitable set of means $\{x_\sigma\}$ and a suitable
number $N_{eff}$ representing the number of groups. A similar argument can be applied also to the MFP model. In this case, each Twitt, besides 
replicating some other Twitts, tends to follow a common global collective mean behavior with all the other Twitts.

The number $N$ might also contain non human-based messages. 
In fact, as is known, large numbers of fake Twitter accounts (robots) can invade the Twitter activities in a tentative to dope the discussions, their main feature being
the fact that they simply replicate the same Twitt many times. However, at this level, we cannot discriminate the two sources of replications, human and non human.  
Nevertheless, the values that we find for the ratio $N'=N/N_{eff}$ provide a rough but simple way for detecting the fraction of replicating Twitts, both human and non human.
Quite interestingly, we find that $N_{eff}$ is one or two orders of magnitude smaller than $N$, which implies that, on overage, there is a overall replication factor of $N'$ order 
$10 \div 10^2$.

\subsection{Electoral events and analysis from MD and MFP}
Below we apply the models MD and MFP to the Twitter data analyzed in Refs. \cite{Brazil,Brazil-2,UK2010,Spain2016,India2014}.

\begin{figure}[htb]
{\includegraphics[height=2.2in]{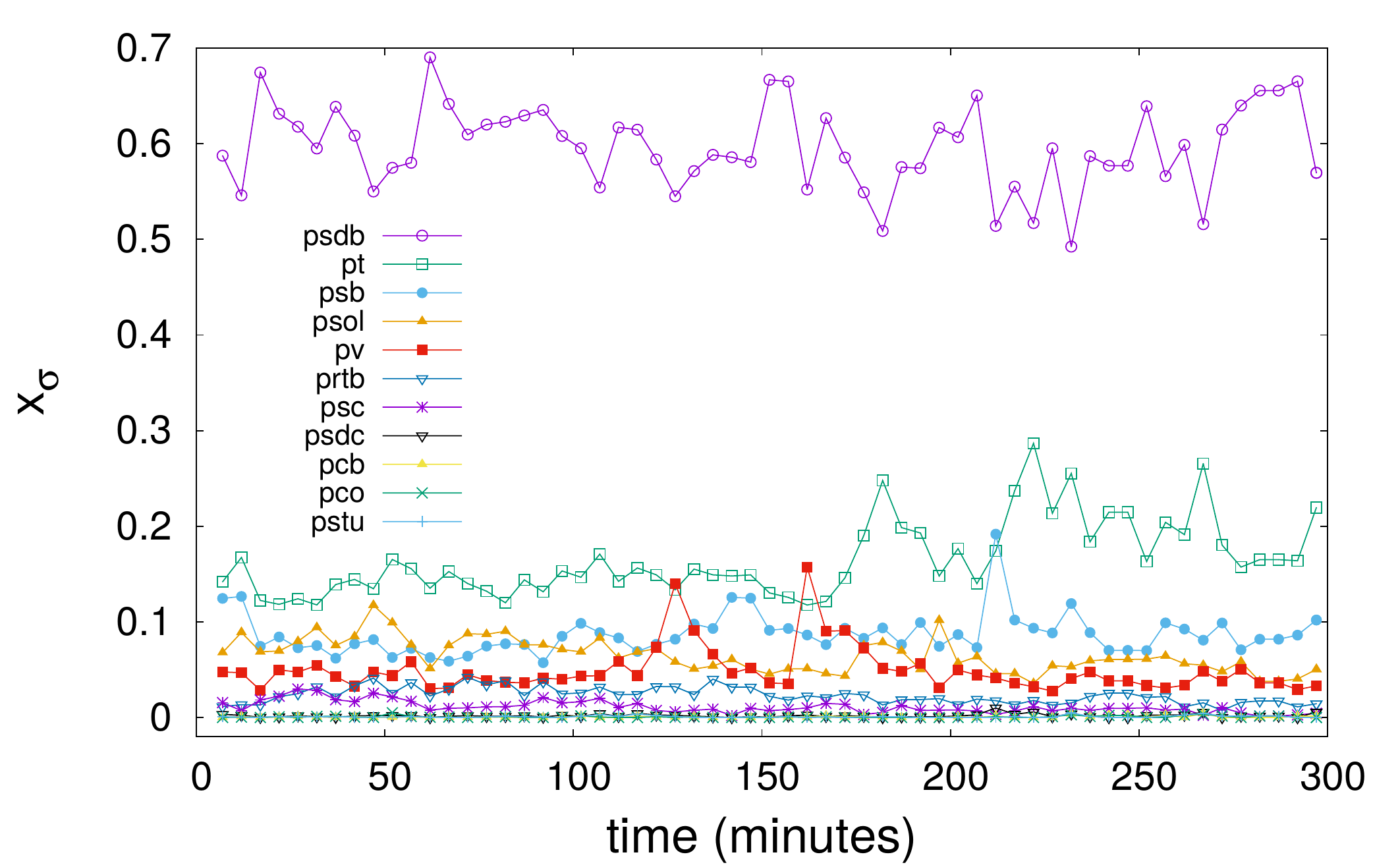}}
\caption{Brazil General Elections 2014, $q=11$: time series of the user mean preferences, or normalized mean number of mentions, $x_\sigma(t)$. 
The acronyms stand for political parties names.}
\label{fig1}
\end{figure}

\begin{figure}[htb]
{\includegraphics[height=2.2in]{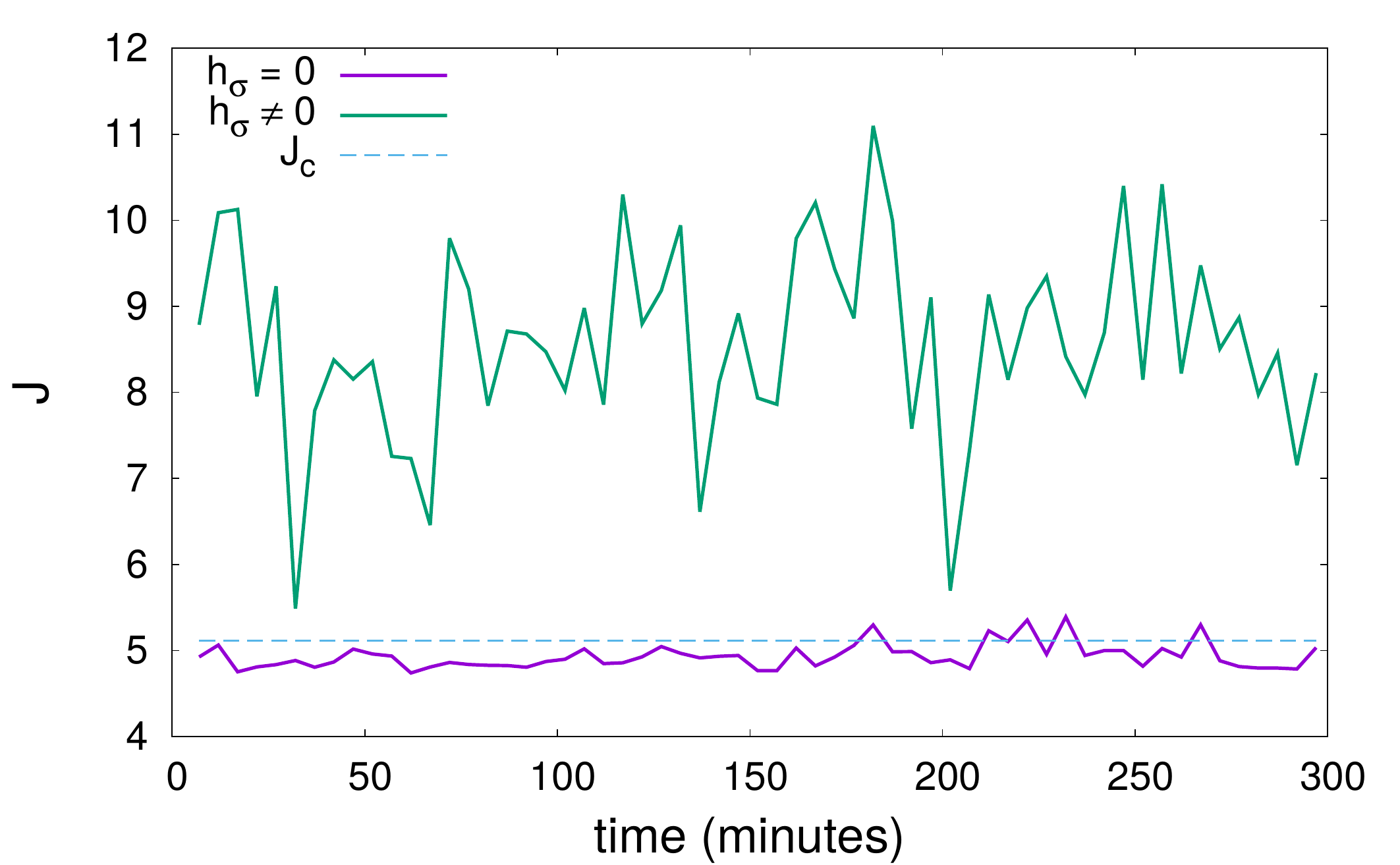}}
\caption{Brazil General Elections 2014, $q=11$: coupling $J(t)$ as solution of the inverse problem with (green) and without (purple) a uniform external field $h_q(t)$.}
\label{fig2}
\end{figure}

\begin{figure}[htb]
{\includegraphics[height=1.4in]{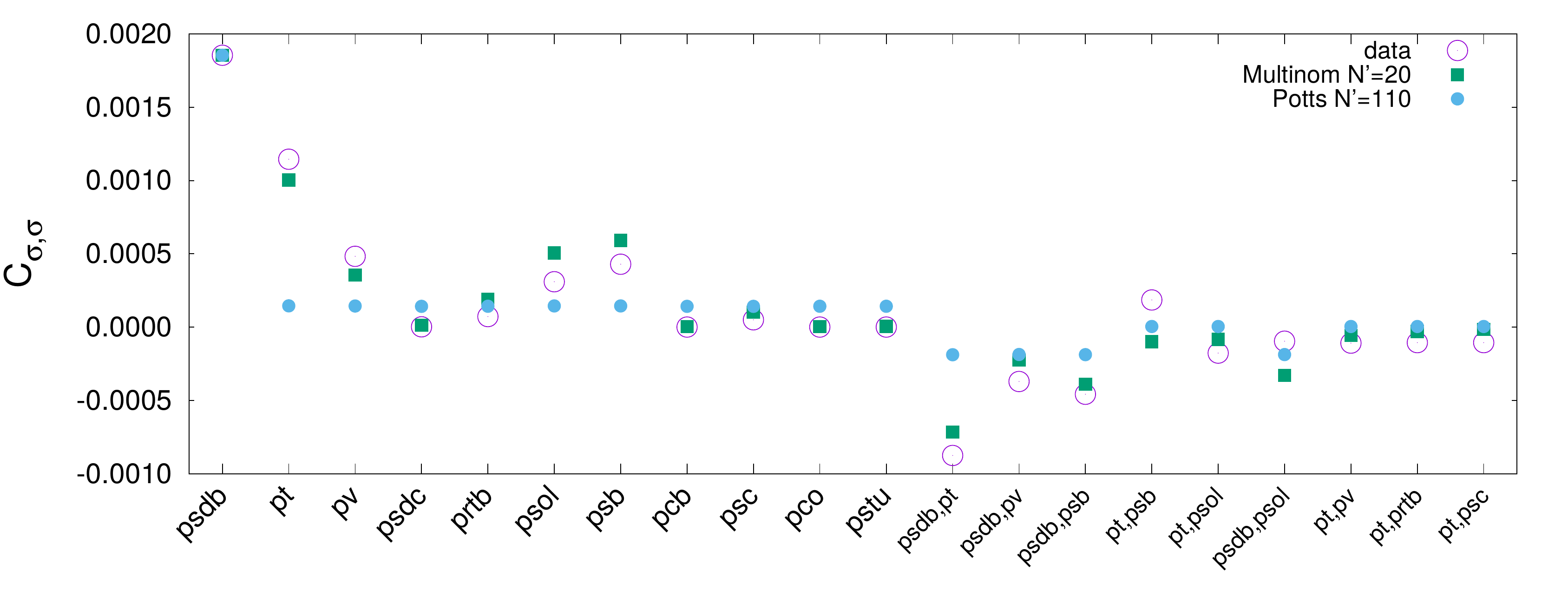}}
\caption{Brazil General Elections 2014, $q=11$: macroscopic correlations from the data and from the MD and MFP models. An acronym in the horizontal axis indicates
one of the $q$ autocorrelations, while a pair of acronyms indicates one of the $q(q-1)/2$ connected correlation functions (for visual convenience, only a few are shown).
For each model, $N'$ represents the factor providing the effective number of messages $N_{eff}=N/N'$ (see Sec. IVC)}
\label{fig3}
\end{figure}

\begin{figure}[htb]
{\includegraphics[height=2.2in]{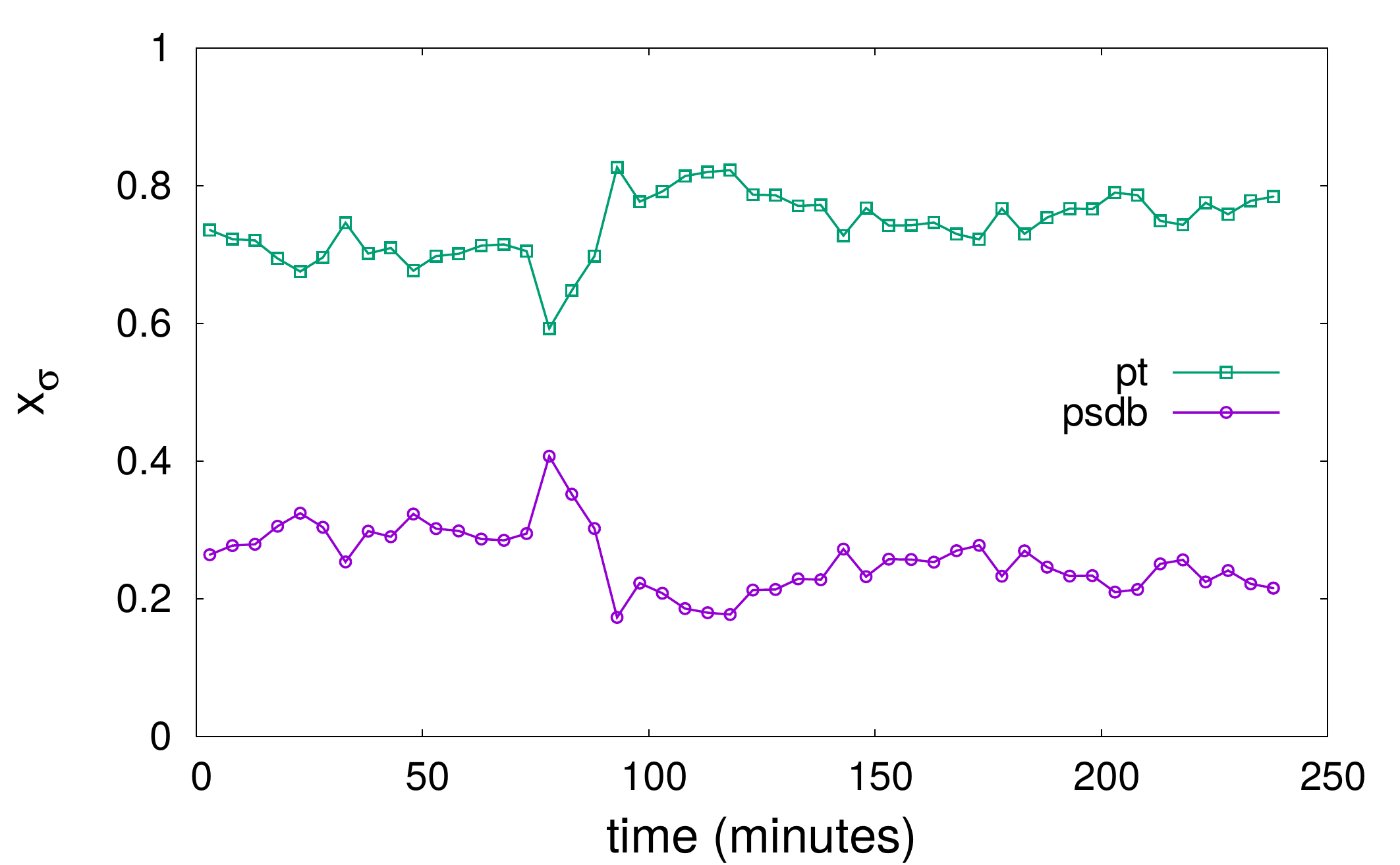}}
\caption{Brazil General Elections 2014 - Second Round, $q=2$: time series of the mean user preferences, or normalized mean number of mentions, $x_\sigma(t)$.}
\label{fig4}
\end{figure}

\begin{figure}[htb]
{\includegraphics[height=2.2in]{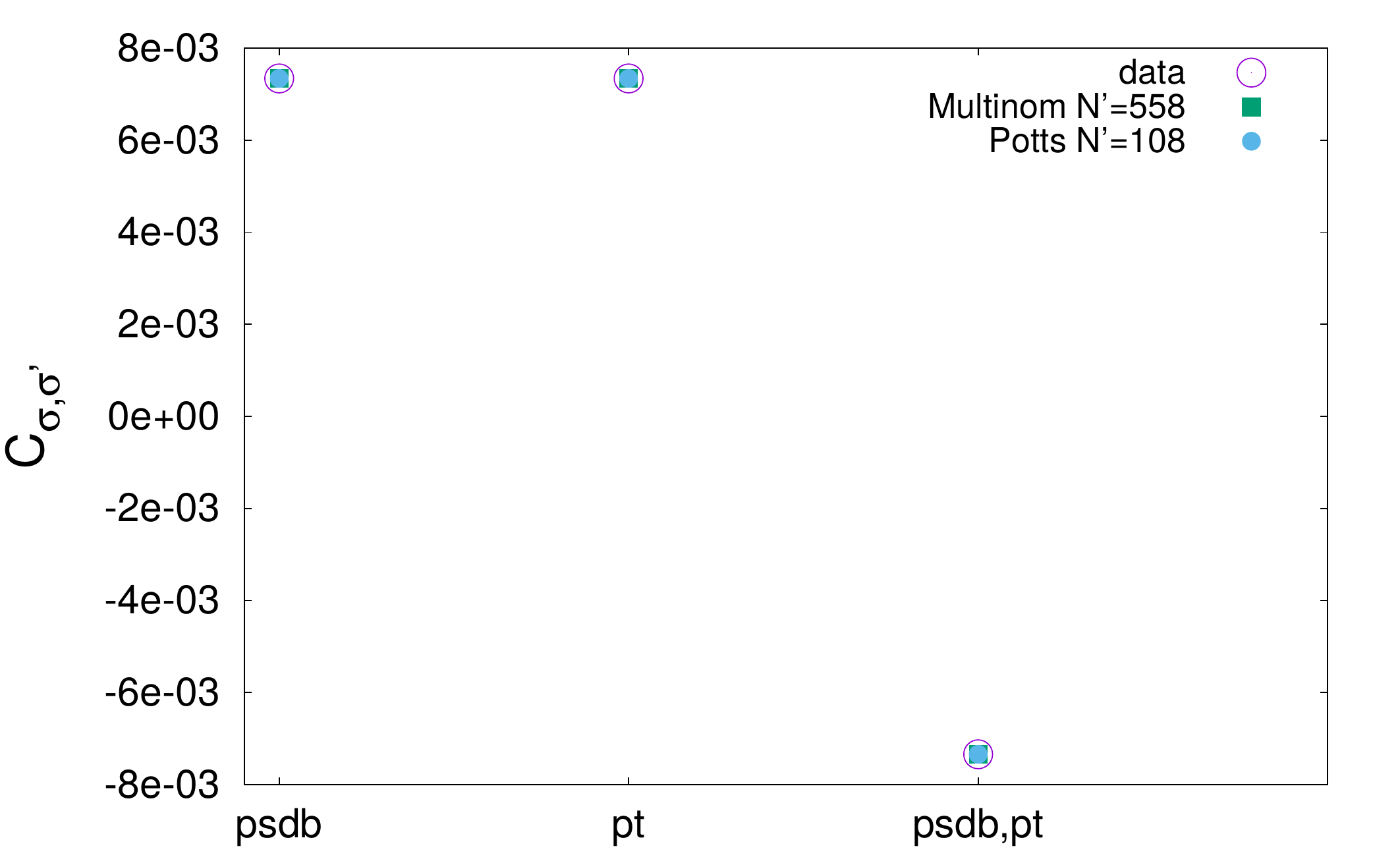}}
\caption{Brazil General Elections 2014 - Second Round, $q=2$: macroscopic correlations (the first two on the left are autocorrelations).
For each model, $N'$ represents the factor providing the effective number of messages $N_{eff}=N/N'$ (see Sec. IVC)}
\label{fig5}
\end{figure}

\begin{figure}[htb]
{\includegraphics[height=2.2in]{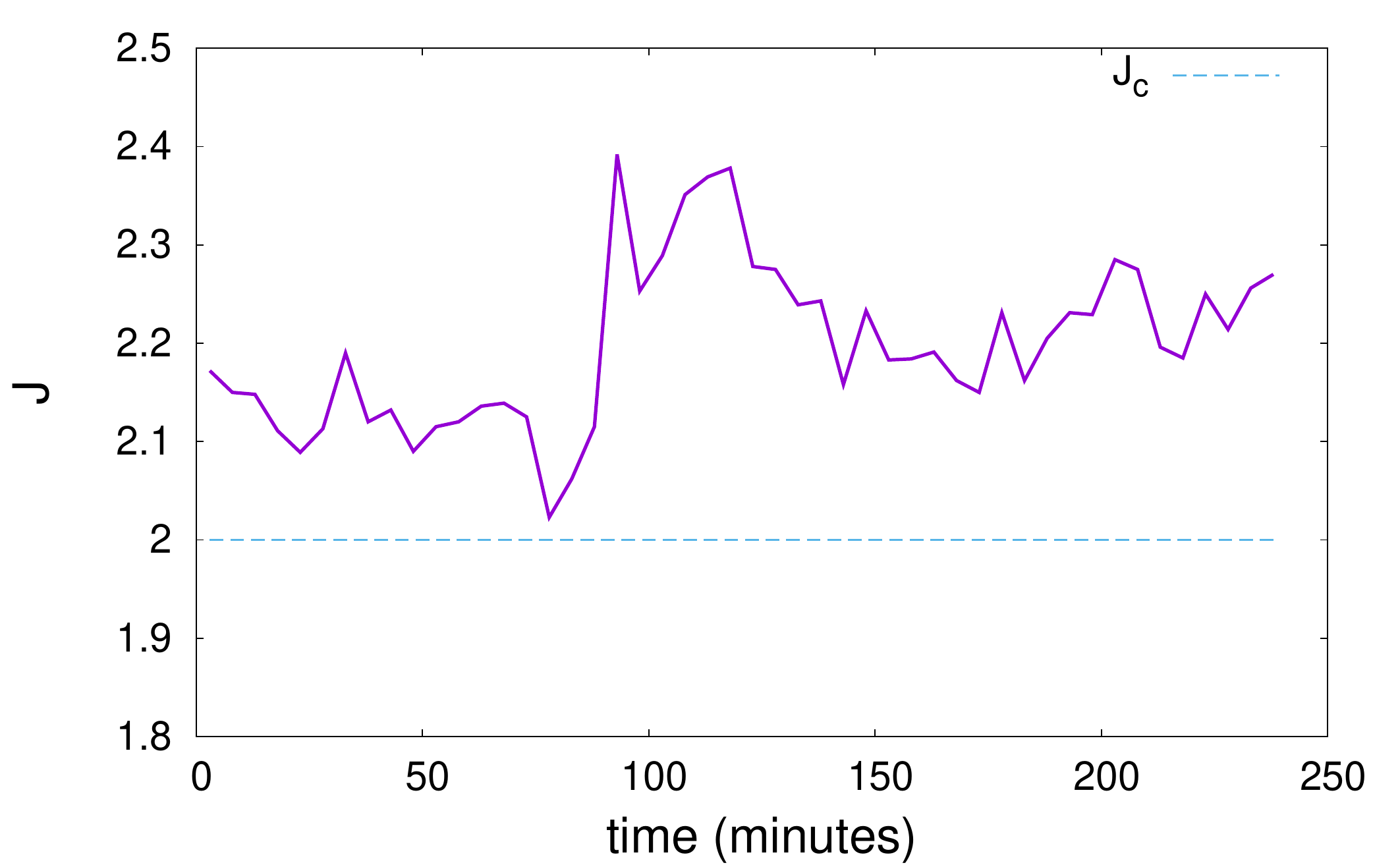}}
\caption{Brazil General Elections 2014 - Second Round, $q=2$:
coupling $J(t)$ as solution of the inverse problem without external field.}
\label{fig6}
\end{figure}

\begin{figure}[htb]
  {\includegraphics[height=2.0in]{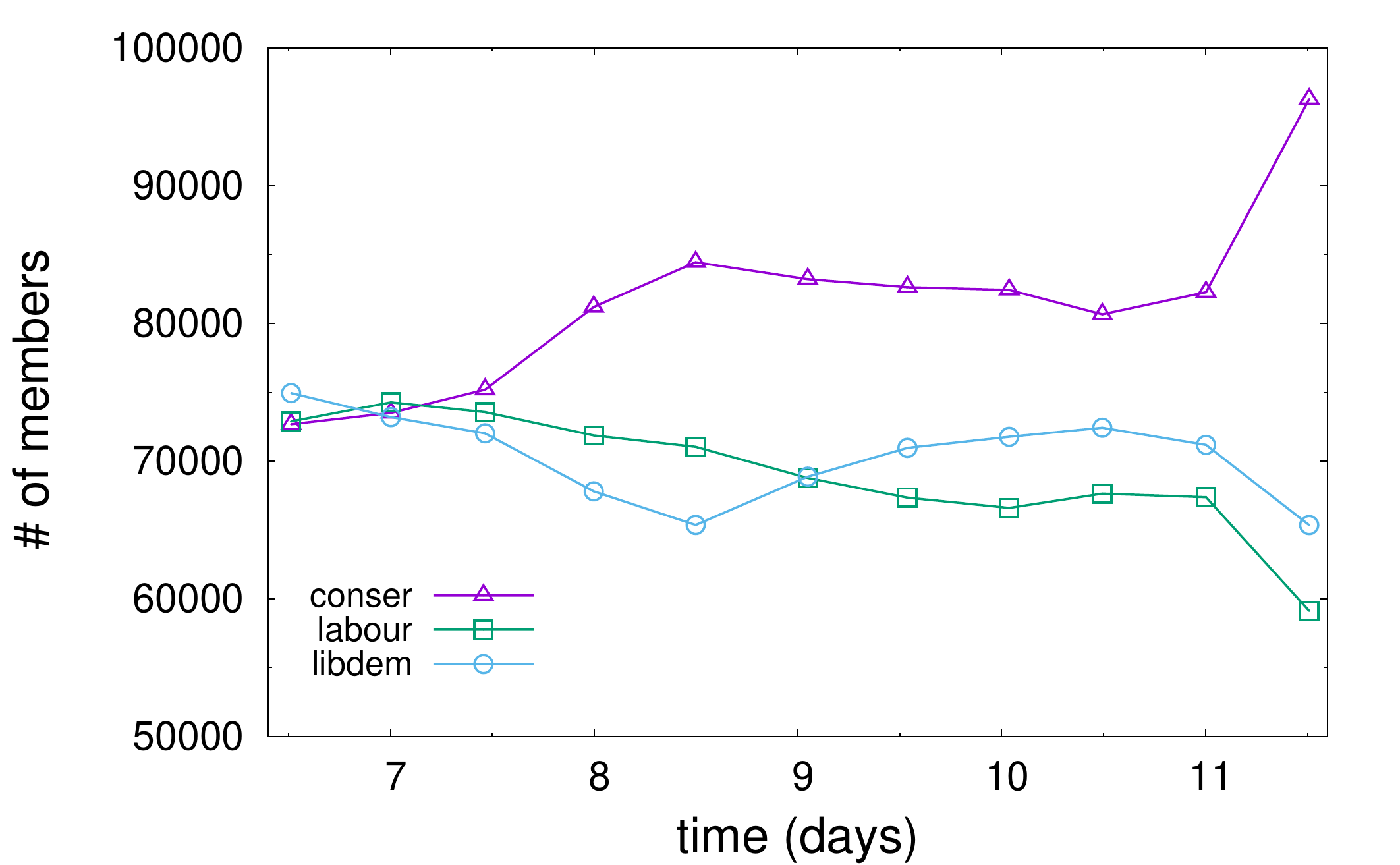}}
\caption{UK General Elections 2010, $q=3$: time series of the mean user preferences, or normalized mean number of mentions, $x_\sigma(t)$.}
\label{fig7}
\end{figure}

\begin{figure}[htb]
  {\includegraphics[height=2.1in]{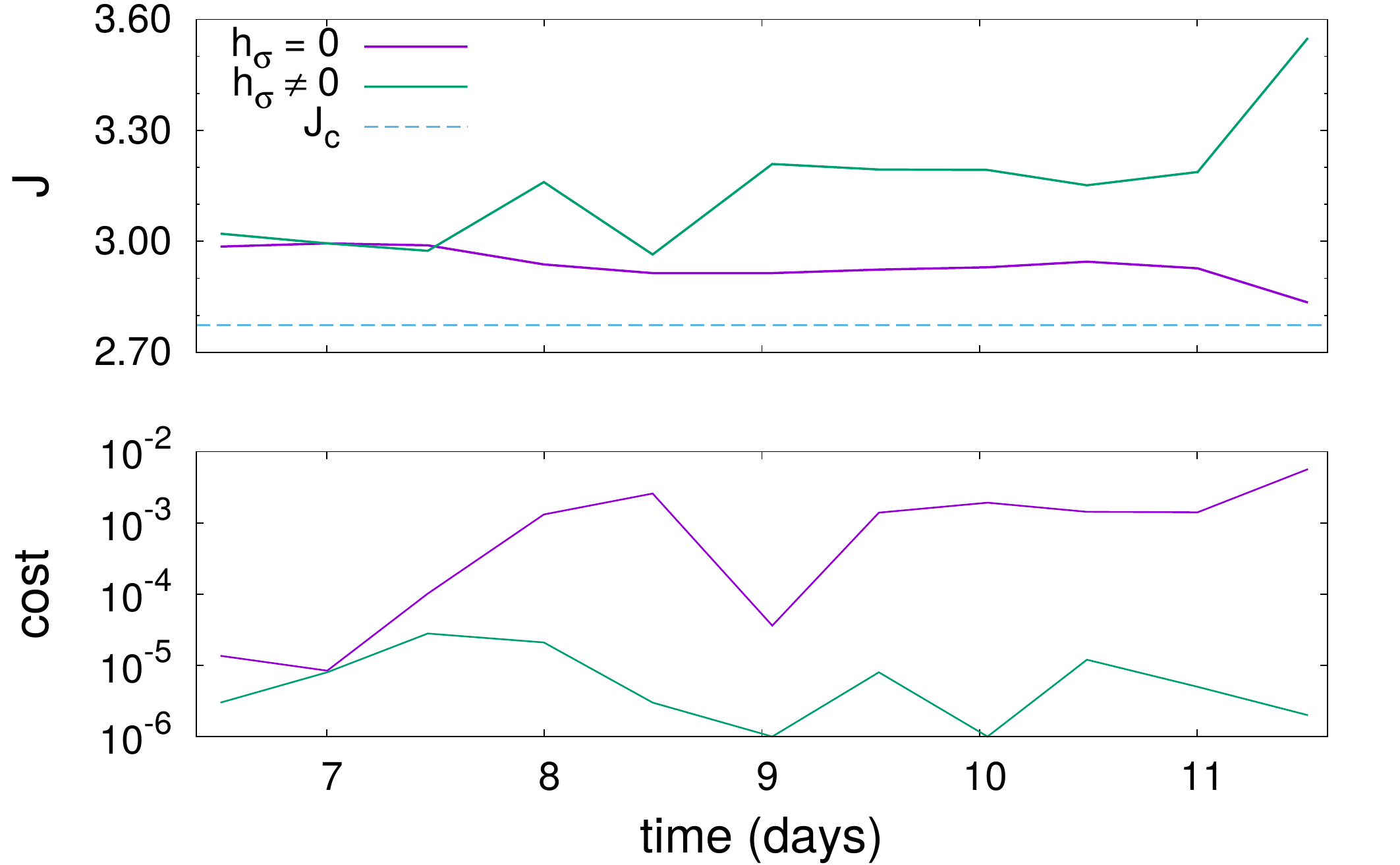}}
\caption{UK General Elections 2010, $q=3$: coupling $J(t)$ as solution of the inverse problem with (green) and without (purple) a uniform external field $h_q(t)$ (upper plot);
residual error of the inverse problem, or cost function (bottom plot).}
\label{fig8}
\end{figure}

\begin{figure}[htb]
{\includegraphics[height=2.2in]{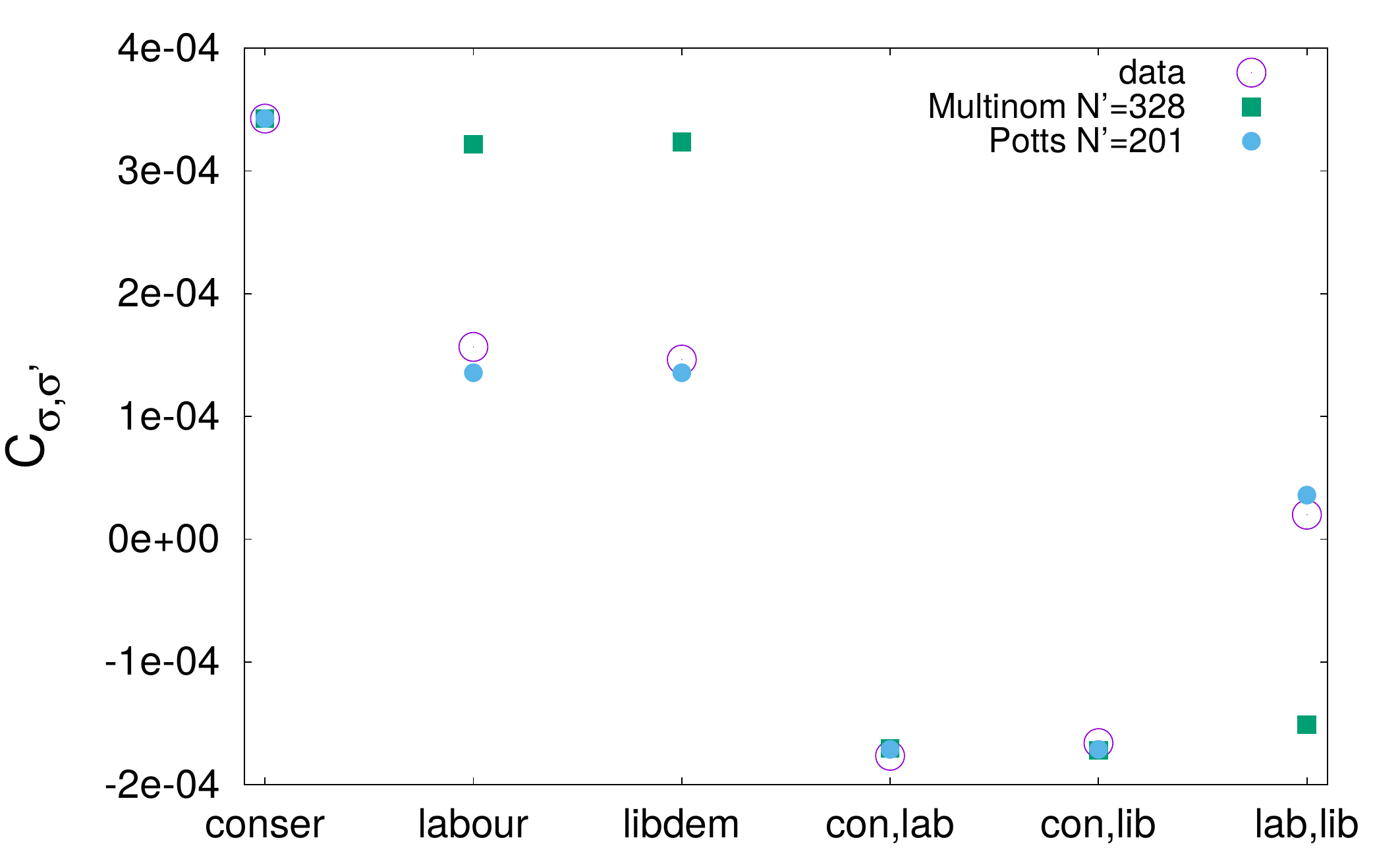}}
\caption{UK General Elections 2010, $q=3$: macroscopic correlations from the data and from the MD and MFP models (the first three on the left are autocorrelations).
For each model, $N'$ represents the factor providing the effective number of messages $N_{eff}=N/N'$ (see Sec. IVC)}
\label{fig9}
\end{figure}

\begin{enumerate}

\item \textbf{Brazil General Elections 2014, $q=11$. \\Source of data: Ref. \cite{Brazil}}\\
This analysis takes into account the 2014 Brazil General Elections with $q=11$ main candidates.
The observation period here is short and the Twitts are followed within a range of 300 minutes around the election day. 
The shortness of the observation period justifies the absence of relevant political news and therefore
the stationary assumption here turns out to be correct. This is also quite evident from the plot of the $x_\sigma(t)$,
where random oscillations seem the only source of changes, see Fig. \ref{fig1}.
The mapping with a MFP with and without an external field is shown in Fig. \ref{fig2}. We see here that, while the matching to a MFP with an external field might be in principle better
(\textit{i.e.}, the error in solving the inverse problem is smaller) with respect to
a case with zero external field, only the latter turns out to be compatible with a stationary regime, since the coupling and external field of the former oscillate violently over time. 
From Fig. \ref{fig2}, we also observe that the coupling $J$ of the MFP without external field turns out quite close to its critical value.
However, by observing in detail the structure of the macroscopic connected correlations in Fig. \ref{fig3}, we find out that the best statistical model coincides with the MD.
This implies that, in this electoral event (first turn), a prediction based on these Twitter data would be legitimate, though, in this election, the winner between the two strongest candidates
remained unclear, and a second round was necessary.  

\item \textbf{Brazil General Elections 2014 - Second Round, $q=2$. \\Source of data: Ref. \cite{Brazil-2}}
The first round of the 2014 Brazil General Elections with $q=11$ did not produce any winner, therefore a second round was afterward performed
between the two first candidates (psdb and pt). 
Also here, taking into account the short period of observation, 
the stationary assumption seems correct, as also evident from the plots of the $x_\sigma(t)$, see Fig. \ref{fig4}.
Due to the fact that here $q=2$, the macroscopic correlations of the MD and MFP models are equal and trivial in this case, see Fig. \ref{fig5}.
As shown in Fig. \ref{fig6}, the solution of the inverse problem for the coupling $J(t) 
$ of MFP model provides a little bit overcritical coupling which leads to understand that a second order phase transition
(for $q=2$ the Potts model is equivalent to the Ising model so that it undergoes second-order phase transitions) 
took place between the end of the first and the beginning of the second round of the election. In fact, it is interesting to observe
that, whereas in the first round the psdb candidate had a net advantage over the pt one, in the second round the order got inverted.
This case, hence, represents a manifested unpredictable example 
(we stress that by prediction here we mean a mere use of trends for making extrapolations; we do not take into account possible
alliances between different parties that could take place between the two rounds).

\item \textbf{UK General Elections 2010, $q=3$. \\Source of data: Ref. \cite{UK2010}}\\
Let us consider Fig. \ref{fig7}. 
It represents the three normalized mean number of mentions $x_\sigma(t)$ collected just after the UK election day, $6$ of May 2010, 
during the period $t=6$ of May till $t=12$ of May. 
This case represents a paradigmatic example: the observation days cover a relatively short period where the news on politics (if any) cannot 
affect much the user's preference toward one or the other candidate because the vote has already been done. 
Therefore, during this period a Twitter user keeps or changes its
preference as a result of the interaction with the other Twitter users without other external influences. 
Furthermore, this electoral event did not produce any winner candidate \textit{per se}. 
In fact, the plot of Fig. \ref{fig7} clearly shows an initial perfect symmetric configuration $x_\sigma=1/q$ followed afterwards by a spontaneous symmetry breaking
taking place during the times $t=6,7,8$ and then evolving toward an approximate stationary configuration with a winner candidate and two almost equal losers.
All this scenario seems therefore well described by a MFP model with 
$q=3$ which, initially, when the configuration is the symmetric one, $x_\sigma=1/q$, is in the paramagnetic phase but out equilibrium,
and then evolves toward equilibrium reaching the polarized solution as given by Eqs. (\ref{UA}-\ref{UA1}). 
Fig. \ref{fig8} shows that the parameters of this MFP model are very close to their critical values as given by Eq. (\ref{tc}) 
(as shown in the bottom plot of Fig. \ref{fig8}, where the residual error of the inverse problem is named ``cost'', 
the matching to a MFP with an external field is in principle better with respect to
a case with zero external field, but only the latter turns out to be compatible with a stationary regime).
This implies that the change from the paramagnetic/symmetric regime toward the ferromagnetic/polarized one, occurs via an abrupt dynamics,
characterized by finite jumps, as also evident from Fig. \ref{fig7}.   
A natural question is whether the MD would be equally able to reproduce the dynamics of Fig. \ref{fig7} if we use the parameters 
$\overline{x}_\sigma$ as given by Eqs. (\ref{MeanData}).
In this sense, one might wonder if the previous description via the MFP is actually a mere coincidence
and the time evolution of the $x_\sigma(t)$ could be equally described via a sequence of random draws from an urn with 3 bias as given by the 
$\overline{x}_\sigma$'s. The analysis of the correlation functions in Fig. \ref{fig9} clearly shows that this is not the case. 
This check hence rules out definitely the MD model in favor of the MFP model,
and since the system is at its critical point, predictions based on these Twitter data would be unreliable. 

\item \textbf{Spanish General Elections 2016, $q=4$. \\Source of data: Ref. \cite{Spain2016}}\\
This analysis takes into account the 2016 Spanish General Elections held on 26 th June, 2016
where the number of main candidates is $q=4$. Here, the observation period corresponds to the days $t=6$ to $t=26$, \textit{i.e.},
the observation period is quite large and prior to the election day. This implies that the stationary assumption can be hardly satisfied here because,
the Twitter users, besides interacting among each other on the base of the Twitter discussions, are also continuously receiving external
news affecting their political preferences. As a consequence, the use of Eqs. (\ref{CorrData}) for measuring the macroscopic correlations
should be avoided. Nevertheless, from the plots of the $x_\sigma(t)$, we observe a kind of approximate stationarity among the user preferences, 
see Fig. \ref{fig10}. Despite these values of $x_\sigma(t)$ seem not well approximated by the MFP model with zero external fields,
and leave the MD model as a more suited one to describe such a dynamics, we find out again that
the coupling and fields that realize the best match between data and MFP are quite close to the critical ones, with $J$ as given 
by Eq. (\ref{tc}) and zero external fields, as shown in Fig. \ref{fig11}. Fig. \ref{fig12} shows also that, 
whereas the matching to a MFP with an external field is in principle better with respect to
a case with zero external field, only the latter turns out to be compatible with a stationary regime.

\item \textbf{Indian Election 2014, $q=2+$sentiment. \\Source of data: Ref. \cite{India2014}}\\
Here, we consider a case with $q=2$ candidates for which the Twitter data provide not only the mentions, but also the sentiment,
defined as the difference between the positive and negative Twitts toward the candidates, where a Twitt is considered
positive or negative according to a dictionary of positive and negative key words and key sentences. 
It is easy to see that this third random variable describing the sentiment, can be treated as a third independent candidate, so that the number of effective
candidates here becomes $q'=3$. As shown by Fig. \ref{fig13}, the model more appropriate here is the MD. This is also evident from Fig. \ref{fig14},
where we see violent oscillations of the coupling of the MFP model, which is not surprising given the large period of observation of 70 days.

\end{enumerate}

\begin{figure}[htb]
{\includegraphics[height=2.2in]{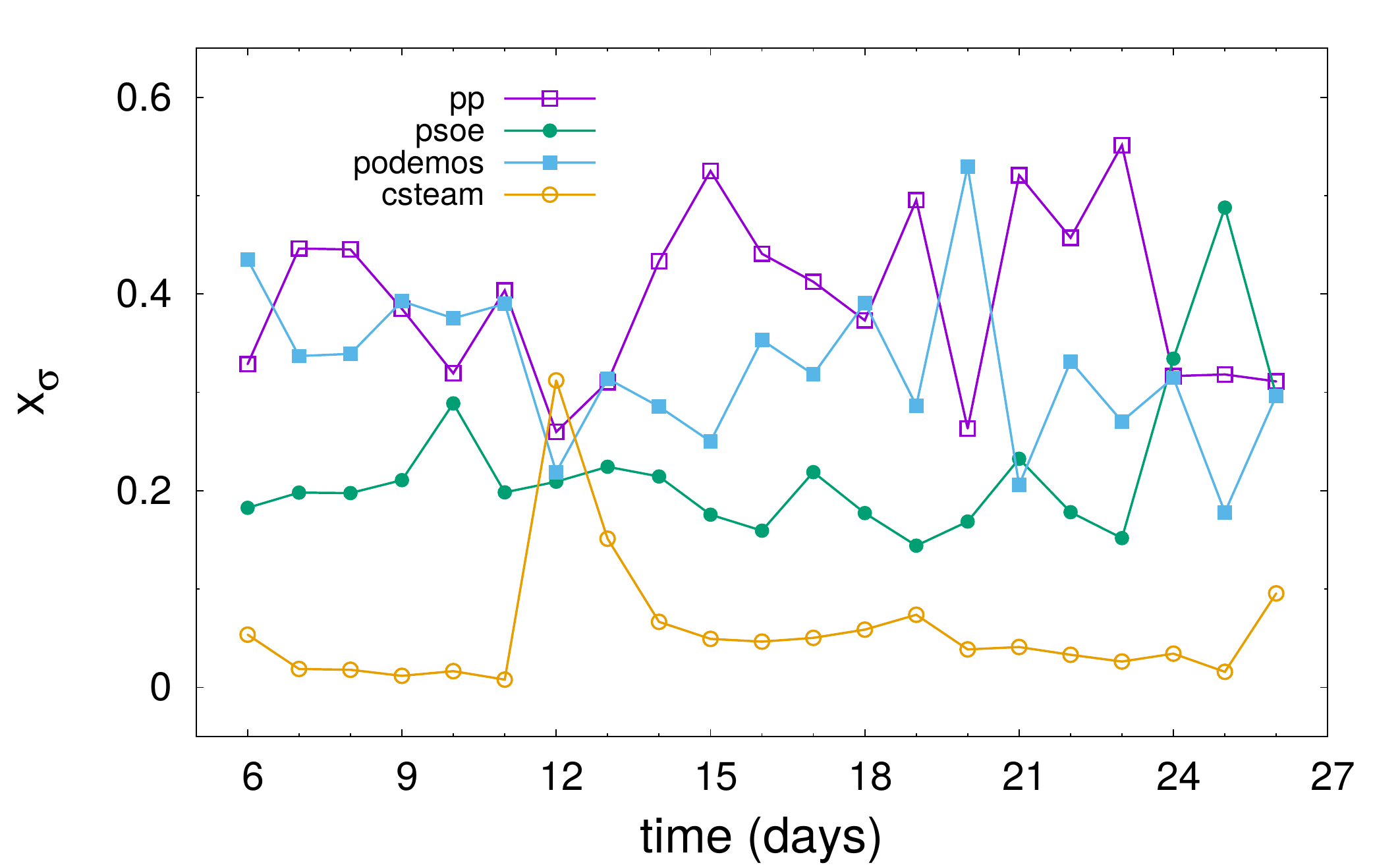}}
\caption{Spanish General Elections 2016, $q=4$: time series of the mean user preferences, or normalized mean number of mentions, $x_\sigma(t)$.}
\label{fig10}
\end{figure}

\begin{figure}[htb]
  {\includegraphics[height=2.2in]{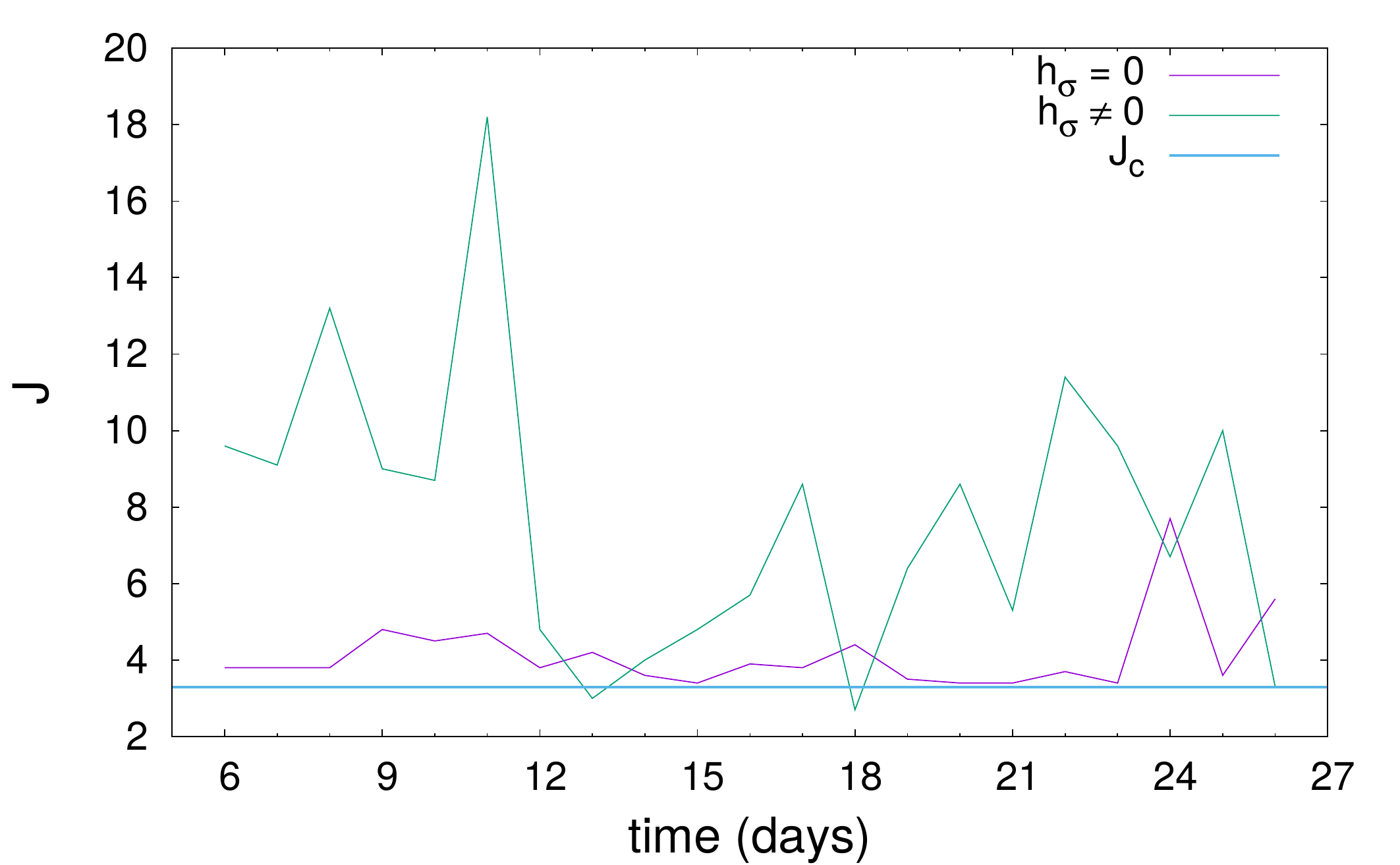}}
\caption{Spanish General Elections 2016, $q=4$: coupling $J(t)$ as solution of the inverse problem with (green) and without (purple) a uniform external field $h_q(t)$.}
\label{fig11}
\end{figure}

\begin{figure}[htb]
{\includegraphics[height=2.2in]{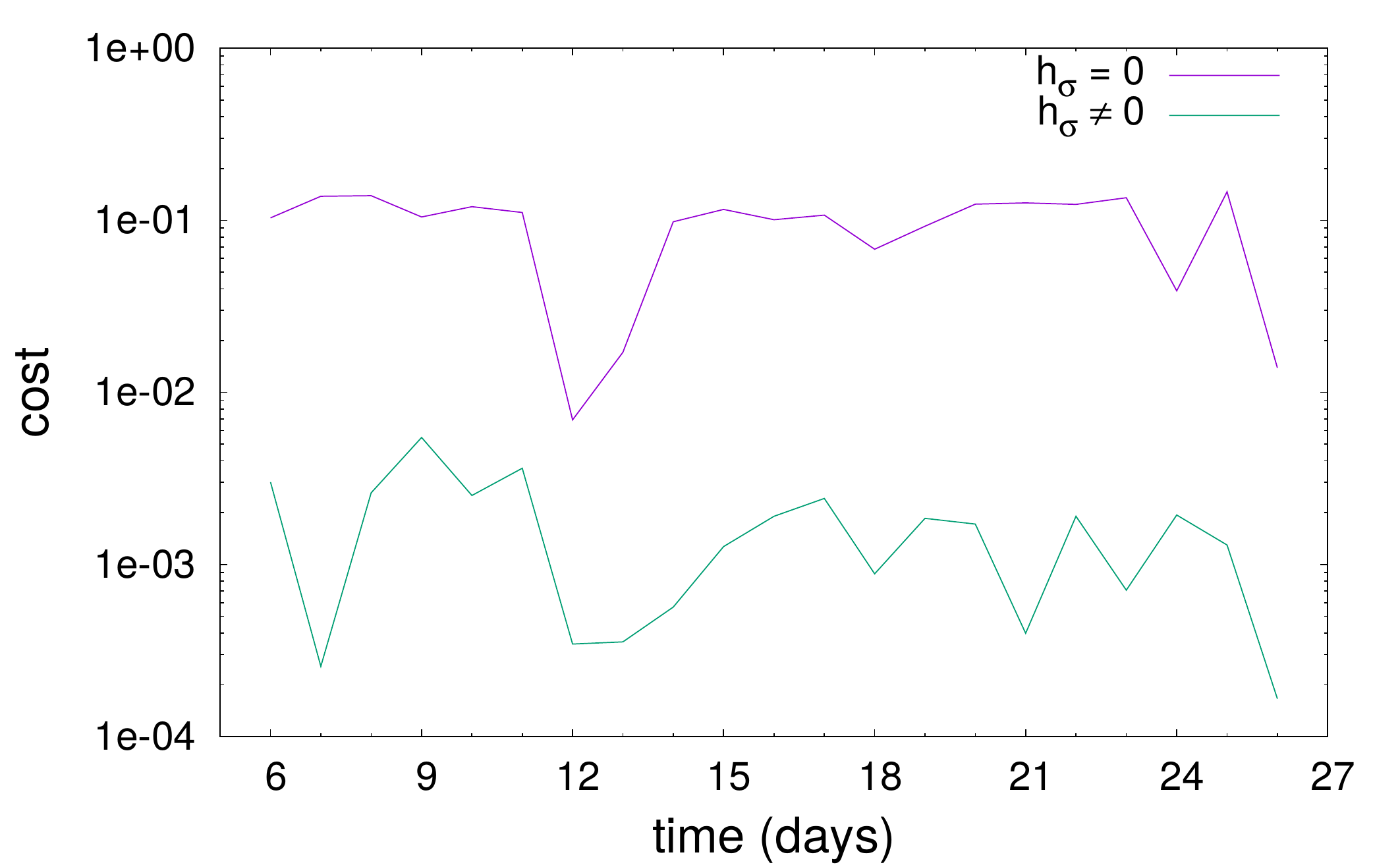}}
\caption{Spanish General Elections 2016, $q=4$: residual error of the inverse problem, or cost function.}
\label{fig12}
\end{figure}

\begin{figure}[htb]
  {\includegraphics[height=2.2in]{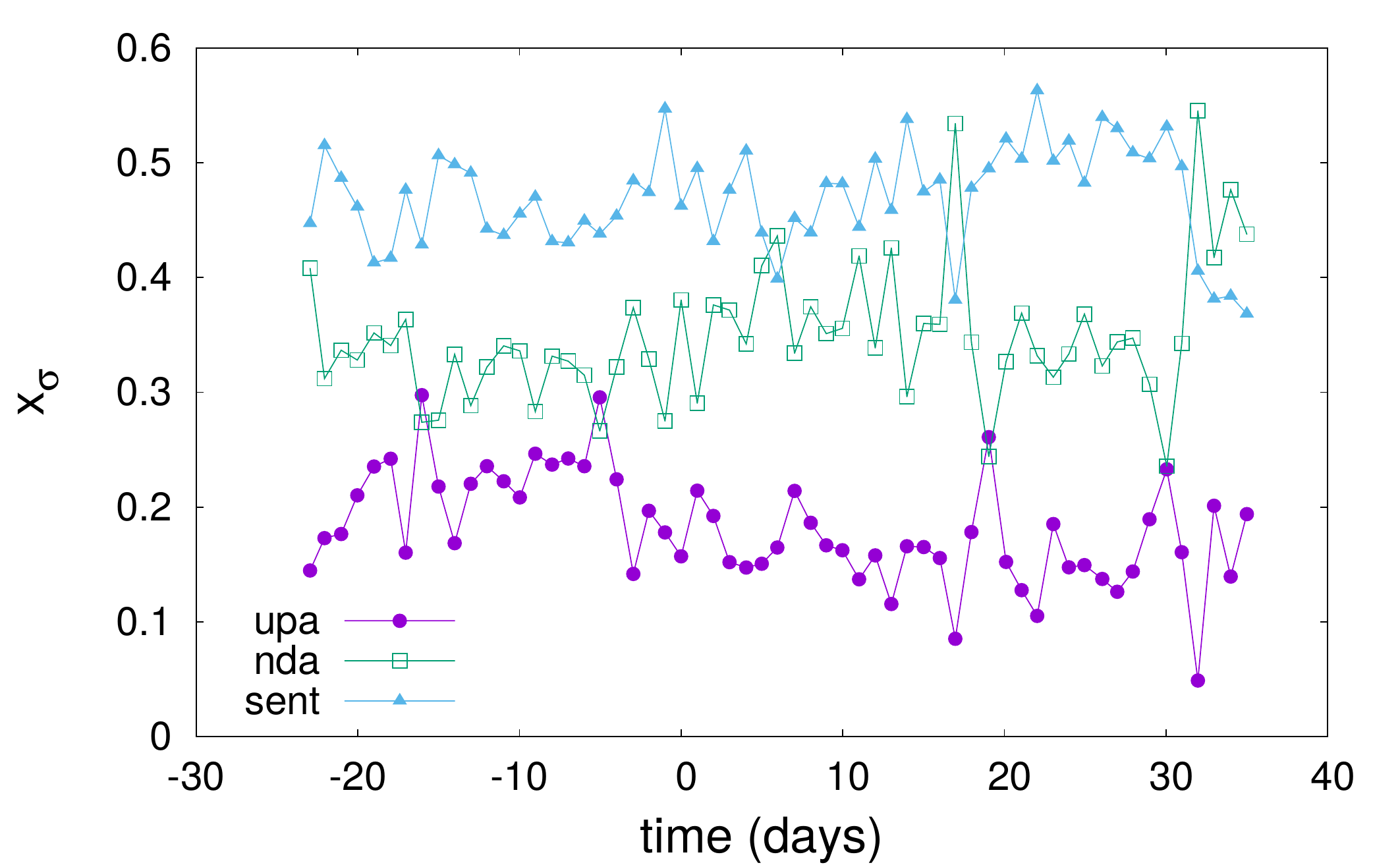}}
\caption{Indian Election 2014, $q=2+$sentiment: time series of two mean user preferences, or normalized mean number of mentions, $x_1(t)$ and $x_2(t)$, and a normalized sentiment 
seen as a further user $x_3(t)$.}
\label{fig13}
\end{figure}

\begin{figure}[htb]
  {\includegraphics[height=2.2in]{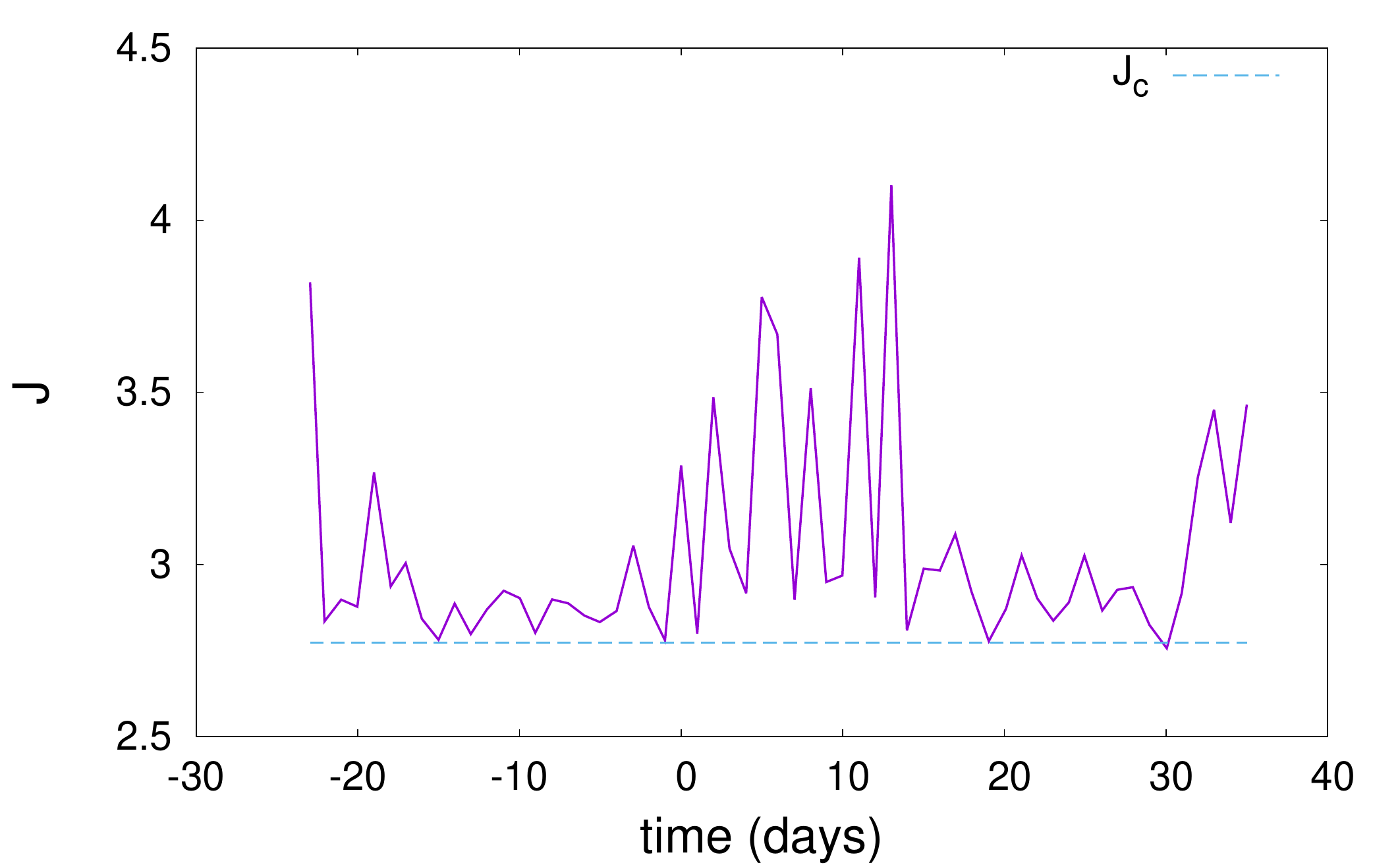}}
\caption{Indian Election, $q=2+$ sentiment: coupling $J(t)$ as solution of the inverse problem without external field.}
\label{fig14}
\end{figure}

\section{Conclusions}
Advanced Big Data technologies can allow to track efficiently the political orientations of a large number of Twitter users (order $10^5$ or more) on the base
of a massive number of Twitts (order $10^6$ or more).
Electoral events provide then an attractive benchmark to test the prediction power of such big polls.  
In the electoral events where a stationary condition applies, and the network of users can be considered as self organized, 
the system of Twitts can be mapped toward two simple models: MD and MFP. In both cases, the correlations decay as $1/N_{eff}$,
$N_{eff}$ being a small fraction of the mean number of Twitts, and the discrimination between the two models can be easily done by looking at
the formulas (\ref{Uncorr4H}) and (\ref{g17}), respectively. 
However, only in the MD case polls can be reliable.  
In the MFP case, instead, the intrinsic symmetry among equally strong candidates, and the empirical observation that these systems lie
close to their critical points, reveals that the winner in this kind of electoral
events is rather the result of a large random fluctuation, preventing then the use of these Twitts for making predictions.
The MD and MFP are the two simplest and somehow opposite effective models able to reproduce the observed statistics. The formal solution of the inverse problem
shows that, richer models, which are generalized Potts models interpolating between the two, could in principle be used for getting
better matchings and further insights. 
In such a desirable plan, however, the parameters involved in these models require the knowledge
of the instantaneous activity of each Twitt, 
which constitute
costly data (in fact, Twitter can release these data on customer demand). 
Furthermore, the solution of the corresponding inverse problem would require a massive analysis.
At any rate, the empirical observation about the critical value of the coupling - a fact that we have observed in all
the electoral events - deserves further investigations.


\section*{Acknowledgment}
We thank Eiko Yoneki for letting us know the Ref. \cite{UK2010}.
Work supported by PNPD/CAPES Grant N. 001/PPGFSC/2017.

\clearpage
\section*{References}


\end{document}